\documentclass[%
 reprint,
superscriptaddress,
showpacs,
 amsmath,amssymb,
 aps,
 prl,
]{revtex4-1}

\usepackage{graphicx}
\usepackage{dcolumn}
\usepackage{bm}
\usepackage{amsmath}
\usepackage{braket}
\usepackage{hyperref}
\usepackage[mathlines]{lineno}

\begin{document}

\preprint{APS/123-QED}

\title{Fundamental precision bounds for three-dimensional optical localization microscopy with Poisson statistics}

\author{Mikael P. Backlund}
\affiliation{Harvard-Smithsonian Center for Astrophysics, Cambridge, MA 02138, USA}
\affiliation{Department of Physics, Harvard University, Cambridge, MA 02138, USA}

\author{Yoav Shechtman}
\affiliation{Department of Biomedical Engineering, Technion, Israel Institute of Technology, Haifa, Israel}

\author{Ronald L. Walsworth}
\affiliation{Harvard-Smithsonian Center for Astrophysics, Cambridge, MA 02138, USA}
\affiliation{Department of Physics, Harvard University, Cambridge, MA 02138, USA}

\date{\today}

\begin{abstract}
Point source localization is a problem of persistent interest in optical imaging. In particular, a number of widely used biological microscopy techniques rely on precise three-dimensional localization of single fluorophores. As emitter depth localization is more challenging than lateral localization, considerable effort has been spent on engineering the response of the microscope in a way that reveals increased depth information. Here we consider the theoretical limits of such approaches by deriving the quantum Cram\'{e}r-Rao bound (QCRB). We show that existing methods for depth localization with single-objective detection exceed the QCRB by a factor $>\sqrt{2}$, and propose an interferometer arrangement that approaches the bound. We also show that for detection with two opposed objectives, established interferometric measurement techniques globally reach the QCRB.
\end{abstract}

\maketitle


Precise spatial localization of single fluorescent emitters is at the heart of a number of important advanced microscopy techniques, including defect-based sensing \cite{RefWorks:doc:5a96d632e4b04ec5e747c2a2,RefWorks:doc:5a96d66ae4b027b722a9bba8,RefWorks:doc:5a96d6a7e4b0642c4190f839,RefWorks:doc:5a96d6f2e4b0642c4190f85e} and single-molecule-based tracking and super-resolution imaging \cite{RefWorks:doc:58d6ed0ce4b05fe42c93d941,RefWorks:doc:59d57285e4b0954cd3d740dd,RefWorks:doc:59d57455e4b07a74c3b17159}. For three-dimensional (3D) imaging and tracking, extracting the emitter's $z$ position (i.e., depth) is an enduring challenge. Microscopists have addressed this by engineering the response of the microscope in ways that improve the attainable depth precision \cite{RefWorks:doc:5a09c164e4b0e70d304d6a26,RefWorks:doc:586d17b2e4b00198145b4b75,RefWorks:doc:59d68600e4b0573d0abb7158,RefWorks:doc:586d17c9e4b0147c58343dfc,RefWorks:doc:59d68668e4b061a5d6147ad6,RefWorks:doc:59d685d3e4b061a5d6147ac2,RefWorks:doc:59d68638e4b00aed3eaf3cf5,RefWorks:doc:59d6885ae4b04214d8e4e5e9,RefWorks:doc:59d6888de4b04214d8e4e5ee,RefWorks:doc:59db9677e4b061a5d61518a3,RefWorks:doc:59db975ce4b00f3d39180384,RefWorks:doc:59db977ae4b0573d0abc0767,RefWorks:doc:59db9698e4b02fc0bca64201}, effectively reducing the associated Cram\'{e}r-Rao bound (CRB) \cite{RefWorks:doc:59d646f8e4b04e4ae184a962}. In this work we address a fundamental question: what is the optimal depth precision that can be attained by any such microscope engineering approach? We derive this measurement-independent limit, the quantum Cram\'{e}r-Rao bound (QCRB) \cite{RefWorks:doc:59d568aee4b08398efcdb1f5}, leading to important new insights for 3D optical localization microscopy, as detailed below.

Throughout this Letter we consider semiclassical photodetection in the limit of Poisson counting statistics \cite{RefWorks:doc:5a4d578de4b0c2a8b20ccff5,RefWorks:doc:59d6b3a0e4b0573d0abb7a2c,RefWorks:doc:59d68d25e4b04e4ae184ba79,RefWorks:doc:59d6a59ce4b0392c41dee5b0,RefWorks:doc:5a96ba60e4b0952b36e61f25}. While this simplified approach ignores bunching and antibunching, it is nonetheless ubiquitous in the fluorescence microscopy literature \cite{RefWorks:doc:59d64167e4b00aed3eaf28ca,RefWorks:doc:59d6414de4b061a5d61462a8,RefWorks:doc:59d64194e4b061a5d61462b5,RefWorks:doc:586d17d9e4b00198145b4b7f,RefWorks:doc:586d17e0e4b00198145b4b94,RefWorks:doc:59d667e2e4b04214d8e4df86,RefWorks:doc:59db977ae4b0573d0abc0767}, as it is relevant to many practical microscopy implementations. For such classically behaving light, the term ``quantum Cram\'{e}r-Rao bound'' is a bit of a misnomer-- a result of the concept's origin in the field of quantum statistical parameter estimation \cite{RefWorks:doc:59d568aee4b08398efcdb1f5}. In fact, it can be derived in the present context with minimal reference to quantum mechanics \cite{RefWorks:doc:59d6a59ce4b0392c41dee5b0}. Thus our work is relevant to a broad class of microscopy techniques in which photon correlations are negligible and justifiably ignored.

In step with the growing attention to precise inference of molecular position, single-molecule microscopists have increasingly adapted the formalisms of statistical parameter estimation \cite{RefWorks:doc:59d64167e4b00aed3eaf28ca,RefWorks:doc:59d6414de4b061a5d61462a8,RefWorks:doc:59d64194e4b061a5d61462b5,RefWorks:doc:586d17d9e4b00198145b4b7f,RefWorks:doc:586d17e0e4b00198145b4b94,RefWorks:doc:59d667e2e4b04214d8e4df86}. In this view, the probability of recording a particular realization of a noisy image array $\mathbf{I}$ conditioned on the underlying source position $ \mathbf{x} = [x_1,x_2,x_3]^\text{T} \equiv [x,y,z]^\text{T}$, is $p(\mathbf{I}|\mathbf{x})$. Related to the CRB is the Fisher information (FI) matrix \cite{RefWorks:doc:59d646f8e4b04e4ae184a962}, with elements given by: 
\begin{equation}\label{eq_cFI}
\mathcal{J}_{ij} = \mathrm{E} \left[ \Big( \partial_{x_i} \log{p(\mathbf{I}|\mathbf{x}) } \Big) \Big( \partial_{x_j} \log{p(\mathbf{I}|\mathbf{x})} \Big) \middle| \mathbf{x} \right],
\end{equation}
where $\mathrm{E}[\cdot|\mathbf{x}]$ denotes the expectation value conditioned on the value of $\mathbf{x}$. The counts $I(x_I,y_I)$ recorded at each position $(x_I,y_I)$ are assumed to be independent and distributed according to $ I(x_I,y_I)|\mathbf{x} \sim \mathrm{Poisson} \left( \bar{I}(x_I,y_I;\mathbf{x}) \right) $ for some expected image $\bar{I}(x_I,y_I;\mathbf{x})$ that depends on the microscope's response function. The same statistics can be obtained from a quantum optical treatment by considering thermal light in the weak-source limit \cite{RefWorks:doc:5a4d578de4b0c2a8b20ccff5,RefWorks:doc:59d6b3a0e4b0573d0abb7a2c,RefWorks:doc:59d68d25e4b04e4ae184ba79}. Equation (\ref{eq_cFI}) then becomes:
\begin{equation}\label{eq_cFI_tot_pois}
\mathcal{J}_{ij} = \iint \mathrm{d}A_I \frac{\left( \partial_{x_i}\bar{I}(x_I,y_I;\mathbf{x}) \right)\left( \partial_{x_j}\bar{I}(x_I,y_I;\mathbf{x}) \right)}{\bar{I}(x_I,y_I;\mathbf{x})}.
\end{equation}
We take the convention that $\bar{I}(x_I,y_I;\mathbf{x})$ is normalized; in accordance with our assumptions of statistical independence then the FI for $N$ detected photons is simply $\mathcal{J}^{(N)} = N \mathcal{J}$. The photon-normalized CRB for the parameter $x_i$ is then given by:
\begin{equation}\label{eq_cCRB}
\sigma^\text{(CRB)}_{x_i} = \sqrt{[\mathcal{J}^{-1}]_{ii}},
\end{equation}
which sets the lower bound for the precision with which any unbiased estimator of $x_i$ can perform \cite{RefWorks:doc:59d646f8e4b04e4ae184a962}.

\begin{figure}[!ht]
\includegraphics[width=8.6cm]{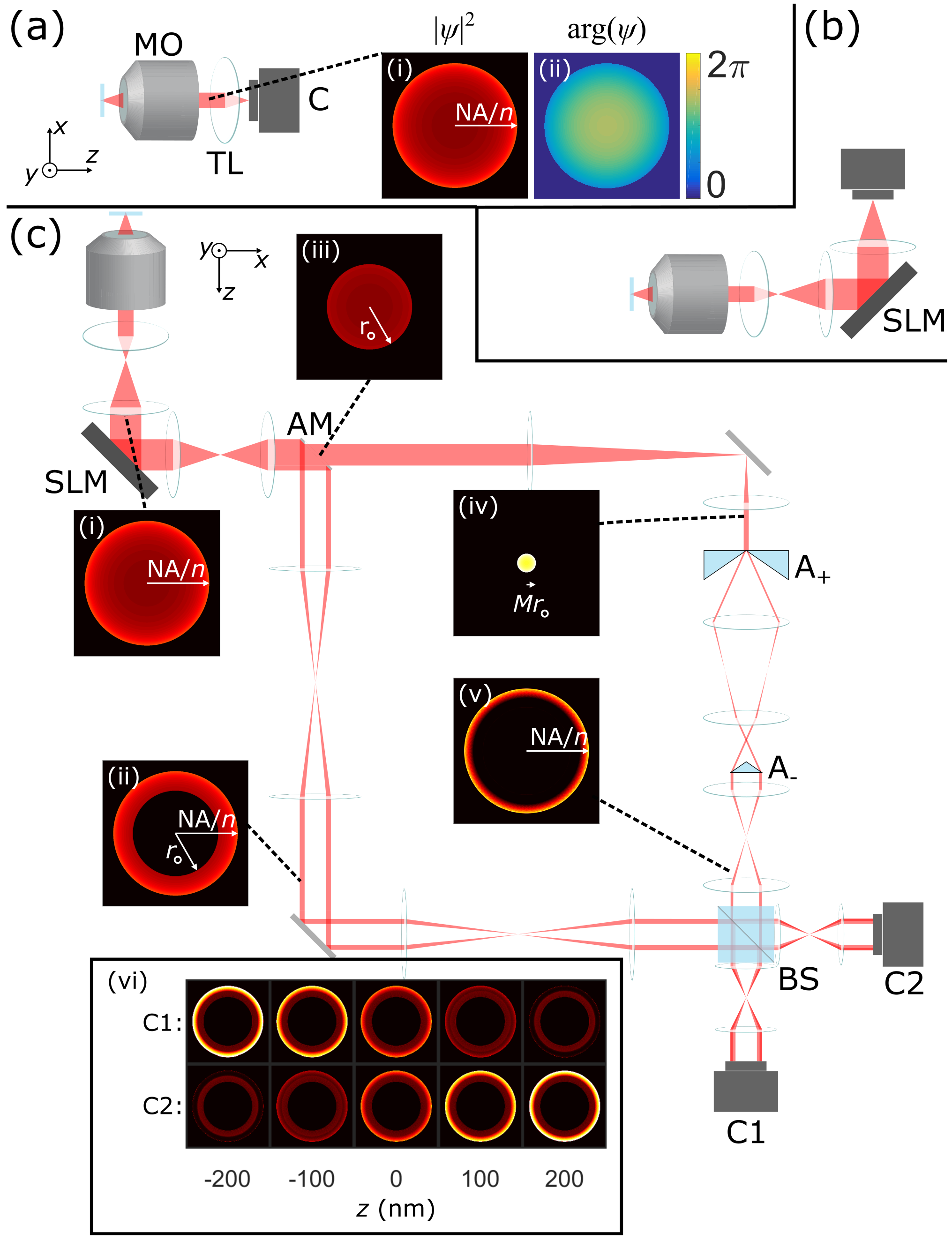}
\caption{Single-objective microscope schematics (emission paths shown). (a) Standard microscope consisting of microscope objective (MO), tube lens (TL) and camera (C). Insets show (i) intensity and (ii) an example phase of $\psi$. (b) Engineered microscope with phase element. An additional pair of lenses can be added to form a 4\textit{f} optical processing unit \cite{RefWorks:doc:58b4a33ce4b07d79a1c92f46}, within which a phase retarder such as a spatial light modulator (SLM) can be placed to change the response function of the microscope. (c) Proposed radial shear interferometer for obtaining $\sigma^{\text{(QCRB)}}_z$. An SLM can be used to compensate for a small amount of defocus imparted downstream. The circular beam (i) is split by an annular mirror (AM) with inner radius $r_{\circ}$. The ``outer'' portion (ii) is relayed to the beam splitter (BS) with two unit-magnification telescopes. The ``inner'' arm (iii) is demagnified with a telescope of chosen magnification $M$ (iv), expanded into a ring of constant thickness with an axicon (A\textsubscript{+}) of phase $\varphi(r_F) = 680 \times r_F$, passed through two relay lenses, collimated with an axicon (A\textsubscript{-}) of phase $\varphi(r_F) = -680 \times r_F$ (v), then relayed to the BS. Average intensities illustrated in insets (i)-(v) have the same color scale, except (iv) which has a $5\times$ scale to avoid saturation. Interferometeric signals are detected on two cameras C1 and C2 placed at conjugate Fourier planes. Example images recorded on C1 and C2 for various $z$ are shown in (vi). The exact distances between optical elements and the diffraction integrals that describe propagation through the apparatus are detailed in \cite{RefWorks:doc:59d6a369e4b04214d8e4ea74}.}
\label{fig_single_obj_sketches}
\end{figure}

We consider a stochastic field with the following normalized equal-time mutual coherence function \cite{RefWorks:doc:5a96ba60e4b0952b36e61f25,RefWorks:doc:59d6a59ce4b0392c41dee5b0,RefWorks:doc:59d68d25e4b04e4ae184ba79,RefWorks:doc:5a4d578de4b0c2a8b20ccff5,RefWorks:doc:59d6b3a0e4b0573d0abb7a2c} on the (Fourier) back focal plane of the microscope objective:
\begin{equation} \label{eq_g_defn}
	g\left(x_F,y_F,x'_F,y'_F;\mathbf{x}\right) = \psi\left(x_F,y_F;\mathbf{x}\right) \psi^*\left(x'_F,y'_F;\mathbf{x}\right).  
\end{equation} 
Here the classical wavefunction in the scalar approximation (in appropriately scaled coordinates) is given by \cite{RefWorks:doc:59d67d4be4b0dbf6a359f412}:
\begin{multline}\label{eq_FP_field}
\psi(x_F,y_F;\mathbf{x}) = \mathcal{A} (1-r_F^2)^{-1/4} \text{Circ}\left(\frac{n r_F}{\mathrm{NA}}\right) \\ \times\exp{\left[ik\left(xx_F + yy_F + z\sqrt{1-r_F^2}\right)\right]},
\end{multline}
as illustrated in Fig. \ref{fig_single_obj_sketches}(a). In Eq. (\ref{eq_FP_field}) $r_F = \sqrt{x_F^2 + y_F^2}$, $n$ is the index of refraction of the objective immersion medium (assumed matched to that of the sample), NA is the numerical aperture, and
\begin{equation}\label{eq_circ_def}
\text{Circ}\left(\frac{n r_F}{\mathrm{NA}}\right) = 
\left\{
	\begin{array}{ll}
		1  & \mbox{if } r_F \leq \frac{\mathrm{NA}}{n} \\
		0 & \mbox{else}
	\end{array}
\right..
\end{equation}
$\mathcal{A}$ is a normalization factor such that $\iint \mathrm{d}A_F|\psi(x_F,y_f)|^2 = 1$, given analytically by:
\begin{equation}\label{eq_A_def}
\mathcal{A} = \left[ 2 \pi \left(1 - \sqrt{1 - (\mathrm{NA}/n)^2} \right) \right]^{-1/2}.
\end{equation}
We assume a quasi-monochromatic signal with free-space wavelength $\lambda_{\circ}$ and $k = 2 \pi n/\lambda_{\circ}$.
After the objective we assume paraxial propagation through air and linear optical elements. We neglect polarization effects, as is appropriate, e.g., for emission from a freely tumbling fluorophore \cite{RefWorks:doc:58d6ee0fe4b05fe42c93d968}. Note that in Eq. (\ref{eq_FP_field}), the source position $\mathbf{x}$ affects only the phase at the Fourier plane, based on the assumption that displacements in $\mathbf{x}$ are sufficiently small \cite{RefWorks:doc:586d1759e4b05e0ea0634282}.
Thus recent work on multi-phase estimation is relevant \cite{RefWorks:doc:5a96bc0de4b025568d7175de,RefWorks:doc:59db8849e4b04214d8e577ff}, though again we stress the classical nature of the problem at hand. In pursuit of the ultimate precision bounds, we here consider the limiting case of zero background light. If energy is conserved between the Fourier plane and the detector, the mean intensity on the camera is related to $\psi(x_F,y_F)$ via a generic unitary operator $U$:
\begin{equation}\label{eq_IUpsi}
\bar{I}(x_I,y_I;\mathbf{x}) = \left| U \circ \psi(x_F,y_F;\mathbf{x}) \right|^2,
\end{equation}
where ``$\circ$'' indicates function composition. Thus once $U$ is specified one can compute Eqs. (\ref{eq_cFI_tot_pois}) and (\ref{eq_cCRB}). The form of $U$ depends on the sequence of optical elements (lenses, mirrors, beam splitters, phase elements, etc.) placed between the Fourier plane and the camera. In the simplest case only a tube lens is added [Fig. \ref{fig_single_obj_sketches}(a)], and the appropriate unitary operation is a scaled Fourier transform $U = \mathcal{F}$ \cite{RefWorks:doc:58b4a33ce4b07d79a1c92f46}. It is known that this approach produces worse FI for $z$ estimation than for $x$ and $y$, especially near $z = 0$ \cite{RefWorks:doc:59d64194e4b061a5d61462b5}.

New microscope designs have been developed in recent years with the goal of modifying the PSF in a way that decreases $\sigma_z^{\text{(CRB)}}$. A common framework is to modulate the phase at the Fourier plane with some carefully chosen phase mask $\varphi(x_F,y_F)$, e.g., programmed onto a spatial light modulator (SLM) [Fig. \ref{fig_single_obj_sketches}(b)] or using a special lens, such that $ U \circ \psi = \mathcal{F}\circ [\psi \times \exp{(i\varphi)}] $ in Eq. (\ref{eq_IUpsi}). This formalism encompasses astigmatic imaging \cite{RefWorks:doc:586d17b2e4b00198145b4b75}, the double-helix microscope \cite{RefWorks:doc:59d68600e4b0573d0abb7158,RefWorks:doc:586d17c9e4b0147c58343dfc}, and the self-bending PSF \cite{RefWorks:doc:59d68668e4b061a5d6147ad6}, among others \cite{RefWorks:doc:59d685d3e4b061a5d6147ac2,RefWorks:doc:59d68638e4b00aed3eaf3cf5}. Related multifocus techniques \cite{RefWorks:doc:5a09c164e4b0e70d304d6a26,RefWorks:doc:59d6885ae4b04214d8e4e5e9,RefWorks:doc:59d6888de4b04214d8e4e5ee} can be represented by a series of beam splitters and phase elements. FI has previously been used as a figure of merit for comparison of these techniques \cite{RefWorks:doc:59d64194e4b061a5d61462b5,RefWorks:doc:586d17d9e4b00198145b4b7f,RefWorks:doc:59d57455e4b07a74c3b17159}. More recently, Shechtman and coworkers demonstrated a rational approach to PSF design by optimizing the mean FI over a specified depth range with respect to a chosen basis for $\varphi(x_F,y_F)$, yielding the saddle-point \cite{RefWorks:doc:586d17d9e4b00198145b4b7f} and tetrapod PSFs \cite{RefWorks:doc:586d17e0e4b00198145b4b94}. This protocol amounts to specifying a form for $U$, then maximizing FI w.r.t. a set of parameters on which $U$ depends. Here we seek a more fundamental approach with the form of $U$ unconstrained. For this we turn to previous work in quantum statistical inference, in which the problem of maximizing FI over all possible positive operator-valued measures (POVMs) has been treated beginning some fifty years ago \cite{RefWorks:doc:59d68c3ce4b03cc71b452e19,RefWorks:doc:59d68c70e4b04e4ae184ba67,RefWorks:doc:59d568aee4b08398efcdb1f5}.

To establish the appropriate notation, suppose the photons collected by the microscope are in the state denoted by the density operator $\rho(\mathbf{x})$. We can then define the quantum Fisher information (QFI) $\mathcal{K}$ associated with this state \cite{RefWorks:doc:59d68c3ce4b03cc71b452e19,RefWorks:doc:59d68c70e4b04e4ae184ba67,RefWorks:doc:59d568aee4b08398efcdb1f5,RefWorks:doc:59d6a92ae4b04e4ae184bf2b,RefWorks:doc:59d6a9afe4b03cc71b45322e}:
\begin{equation}\label{eq_QFI_def}
\mathcal{K}_{ij} = \frac{1}{2} \text{Re } \text{Tr } \rho \left( \mathcal{L}_{x_i}\mathcal{L}_{x_j} + \mathcal{L}_{x_j}\mathcal{L}_{x_i} \right),
\end{equation}
where $\mathcal{L}_{x_i}$ is the symmetric logarithmic derivative defined implicitly by:
\begin{equation}\label{eq_SLD_def}
\partial_{x_i} \rho = \frac{1}{2} \left( \mathcal{L}_{x_i} \rho + \rho \mathcal{L}_{x_i} \right).
\end{equation}
Analogous to the relation between the CRB and FI, the QCRB is related to the QFI by:
\begin{equation}\label{eq_qcrb_defn}
\sigma^\text{(QCRB)}_{x_i} = \sqrt{[\mathcal{K}^{-1}]_{ii}}.
\end{equation}
The QCRB defined in Eq. (\ref{eq_qcrb_defn}) bounds the estimation precision for any measurement on the state $\rho(\mathbf{x})$ \cite{RefWorks:doc:59d568aee4b08398efcdb1f5}. For our purposes, we have $\sigma_{x_i}^{\text{(CRB)}} \geq \sigma_{x_i}^{\text{(QCRB)}}$, regardless of the microscope configuration after the objective lens. Thus we can compare $\sigma_z^{\text{(CRB)}}$ associated with state-of-the-art techniques to the ultimate bound set by $\sigma_z^{\text{(QCRB)}}$.

To proceed in computing the QFI and QCRB, we specify the single-photon state represented by:
\begin{multline} \label{eq_density_def}
\rho(\mathbf{x}) = \iint \mathrm{d}A_F \iint \mathrm{d}A'_F g(x_F,y_F,x'_F,y'_F;\mathbf{x}) \\ \times \ket{x_F,y_F} \bra{x'_F,y'_F}
\end{multline}
where $\ket{x_F,y_F} = a^\dagger(x_F,y_F) \ket{0}$, and $a^\dagger(x_F,y_F)$ is the creation operator for the specified mode, obeying the canonical commutation relation $[a(x^{\prime}_F,y^{\prime}_F),a^{\dagger}(x_F,y_F)] = \delta(x^{\prime}_F-x_F)\delta(y^{\prime}_F-y_F)$. It should be emphasized that the classical statistical optical state we consider in this work is certainly not equivalent to the highly quantum mechanical one-photon state of Eq. (\ref{eq_density_def}). Rather it can be shown that under the appropriate approximations (thermal light in the weak-source limit), the maximum value of $\mathcal{J}_{ij}$ described in Eq. (\ref{eq_cFI_tot_pois}) is mathematically equivalent to $\mathcal{K}_{ij}$ obtained by substitution of Eq. (\ref{eq_density_def}) in Eq. (\ref{eq_QFI_def}) \cite{RefWorks:doc:59d6a59ce4b0392c41dee5b0,RefWorks:doc:59d68d25e4b04e4ae184ba79}. We adopt a similar strategy to that recently used to examine the related problem of resolving two weak thermal point sources \cite{RefWorks:doc:59d68d25e4b04e4ae184ba79} (which has since inspired a number of theoretical and experimental follow-up studies \cite{RefWorks:doc:59d788d0e4b0573d0abb92f2,RefWorks:doc:59d788a4e4b03cc71b455654,RefWorks:doc:59d78880e4b03cc71b45564f,RefWorks:doc:59d78d46e4b0573d0abb9352,RefWorks:doc:59d78833e4b00f3d39178ac9,RefWorks:doc:59d786c1e4b061a5d614b825,RefWorks:doc:59d78724e4b0dbf6a35a2153,RefWorks:doc:59d78987e4b061a5d614b87d,RefWorks:doc:59d78a88e4b00aed3eaf7730,RefWorks:doc:5a96bb34e4b01d55d0b35be4,RefWorks:doc:59d78afee4b0dbf6a35a21ab,RefWorks:doc:59d78b2fe4b00aed3eaf773f}). The problem of establishing quantum bounds of localizing a single point source has also been considered in a number of contexts over the years \cite{RefWorks:doc:59d68c70e4b04e4ae184ba67,RefWorks:doc:59d568aee4b08398efcdb1f5,RefWorks:doc:59d79217e4b04e4ae184e15d}. We distinguish our work by deriving expressions that lend themselves to direct comparisons to existing 3D localization microscopes.

\begin{figure}[!ht]
\includegraphics[width=8.6cm]{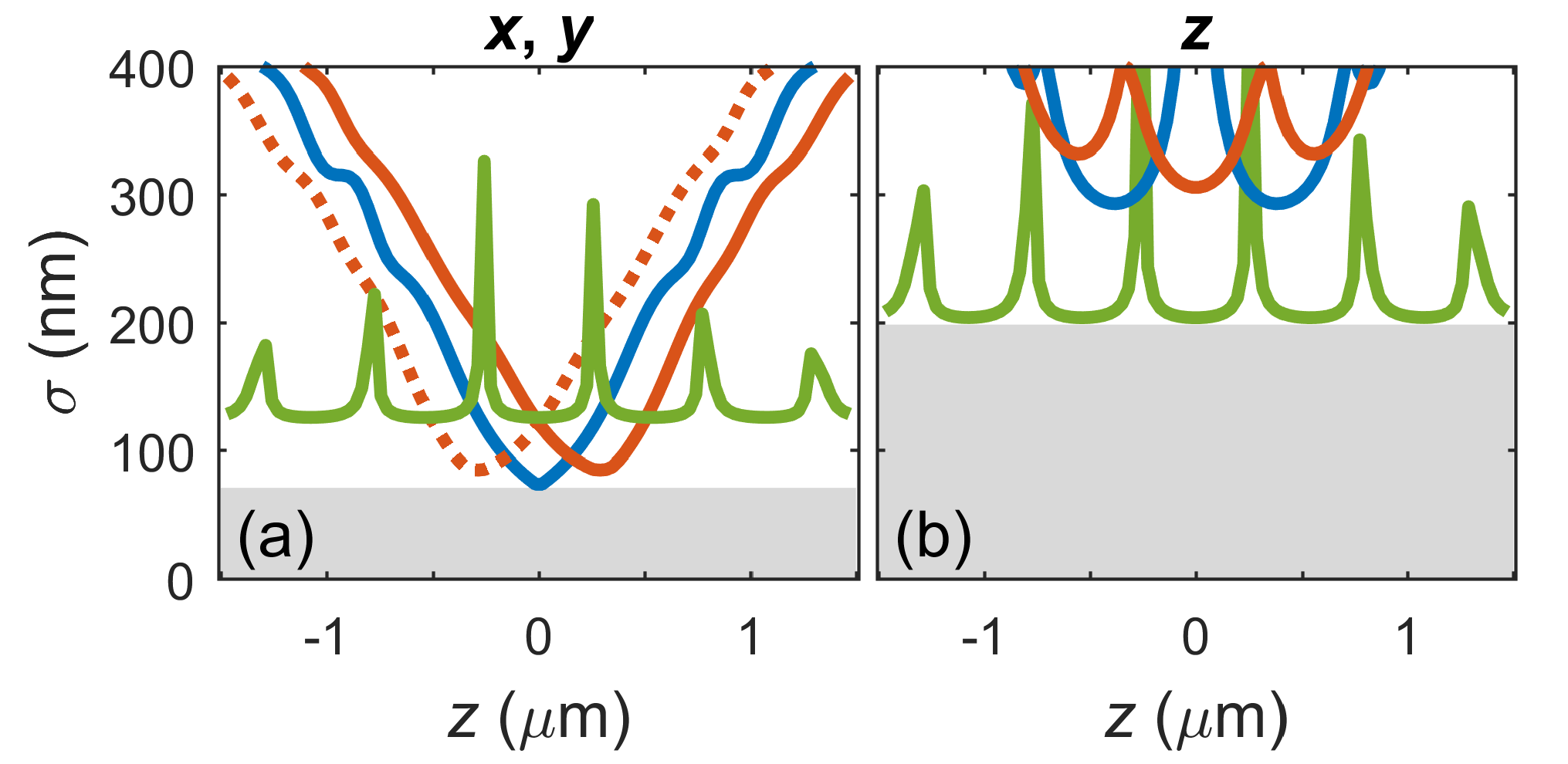}
\caption{Photon-normalized QCRBs and measurement CRBs for single-objective detection as a function of source distance from focus. For $N$ detected signal photons divide vertical axis by $\sqrt{N}$. (a) Lateral localization bounds. The gray shaded region is bounded above by $\sigma = \sigma^{\text{(QCRB)}}_x = \sigma^{\text{(QCRB)}}_y$. Blue curve shows lateral CRB for the standard microscope, red curves (solid is $\sigma^{\text{(CRB)}}_x$ and dotted is $\sigma^{\text{(CRB)}}_y$) show those of astigmatic imaging with strength specified in the main text, green curves shows that of proposed radial shear interferometer. (b) Depth localization bounds. Color code corresponds to that in (a).}   
\label{fig_sigma_single_obj}
\end{figure}

In the Supplemental Material \cite{RefWorks:doc:59d6a369e4b04214d8e4ea74} we derive the QCRBs for 3D localization microscopy using a single objective. The results are:
\begin{subequations}\label{eq_QCRB_3D_singleobjective}
	\begin{align}
    	\sigma_x^{\text{(QCRB)}} &= \sigma_y^{\text{(QCRB)}} = C_{xy}/2, \\ \sigma_z^{\text{(QCRB)}} &= \left( C_z^{-2} - |\gamma|^2 \right)^{-1/2}/2,
    \end{align}
\end{subequations}
with 
\begin{equation}\label{eq_Cxy}
C_{xy} = \frac{\sqrt{3}}{k \mathcal{A}\sqrt{\pi}} \left[ 2 - \sqrt{1 - \left(\mathrm{NA}/n\right)^2} \left( 2 + \left(\mathrm{NA}/n\right)^2 \right) \right]^{-1/2},
\end{equation}
\begin{equation}\label{eq_Cz}
C_z = \frac{\sqrt{3}}{k\mathcal{A}\sqrt{2 \pi}} \left[ 1 - \left(1 - \left(\mathrm{NA}/n \right)^2 \right)^{3/2} \right]^{-1/2},
\end{equation}
and
\begin{equation}\label{eq_gamma}
\gamma = i k \mathcal{A}^2 \pi (\mathrm{NA}/n)^2.
\end{equation}
In Fig. \ref{fig_sigma_single_obj} we compare the QCRBs (gray shaded regions) to the CRBs pertaining to several choices of microscope configuration, with $\mathrm{NA} = 1.4$, $n = 1.518$ (oil immersion), and $\lambda_{\circ} = 670$ nm. The blue lines correspond to a standard microscope configuration [Fig. \ref{fig_single_obj_sketches}(a)]. The red lines show results for astigmatic imaging with $\varphi(x_F,y_F) = \sqrt{6}\left( x_F^2 - y_F^2 \right)$. Here astigmatic imaging of this strength stands in as a representative for similarly engineered PSFs [Fig. \ref{fig_single_obj_sketches}(b)], as justified by the facts that this choice obtains the minimum $\sigma^{\text{(CRB)}}_z$ near $z=0$ for any astigmatic strength and that its local minimum compares favorably to those of other engineered PSFs (Figs. S1 and S2 \cite{RefWorks:doc:59d6a369e4b04214d8e4ea74}). Unsurprisingly, the standard microscope obtains the QCRB for lateral localization precision at focus. However, both the standard and astigmatic configurations exceed the ultimate depth precision limit by a factor $>\sqrt{2}$.

Computing the QCRB is both straightforward and useful, as it gives crucial context for PSF optimization techniques \cite{RefWorks:doc:586d17d9e4b00198145b4b7f}. Establishing conditions for a measurement that attains the bound is a related topic of interest \cite{RefWorks:doc:59d6a9afe4b03cc71b45322e,RefWorks:doc:59db836ce4b061a5d61513ae,RefWorks:doc:59db838de4b04e4ae1854dc1,RefWorks:doc:59db88ece4b00f3d3917fe68,RefWorks:doc:59db88c9e4b03cc71b45d83a,RefWorks:doc:59db8892e4b04214d8e57811,RefWorks:doc:59db87e9e4b00f3d3917fe45,RefWorks:doc:59db8809e4b00f3d3917fe4a,RefWorks:doc:59db8849e4b04214d8e577ff,RefWorks:doc:59db886de4b0dbf6a35a90d3,RefWorks:doc:5a96bc0de4b025568d7175de}. While 3D localization is inherently a multiparameter estimation problem, we here focus on finding a measurement that optimizes the CRB for $z$ estimation, as this is the parameter of primary interest in this work. A sufficient condition for approaching the QCRB for a single parameter is to project onto the eigenstates of the associated SLD \cite{RefWorks:doc:59d6a9afe4b03cc71b45322e}. To approximate projection onto the eigenstates of $\mathcal{L}_z$ (see Fig. S3 and related text in \cite{RefWorks:doc:59d6a369e4b04214d8e4ea74}) we propose the microscope configuration depicted in Fig. \ref{fig_single_obj_sketches}(c), a variant of a radial shearing interferometer \cite{RefWorks:doc:59db8f6ee4b04214d8e57987}. The collected light is split into two parts using an annular mirror \cite{RefWorks:doc:59db9222e4b0573d0abc062b}: an inner disk with support $r_F \leq r_{\circ}=0.6326$, and an outer ring with support $r_F \in (r_{\circ}=0.6326,\mathrm{NA}/n)$. In the ``outer'' arm we (de)magnify the beam by a factor $M=0.22$, then stretch with a pair of axicon prisms \cite{RefWorks:doc:59db8f6ee4b04214d8e57987,RefWorks:doc:59db8fc9e4b00aed3eaff1ab,RefWorks:doc:59db8fefe4b03cc71b45d9cf}. The two portions are recombined with a 50/50 beam splitter and the signal is detected with two cameras placed at conjugate Fourier planes. Some calculated average images are shown in the inset of Fig. \ref{fig_single_obj_sketches}(c) for various $z$. We treat the field classically throughout this work, e.g., neglecting contributions from field operators of modes in the vacuum state at the input of the beam splitter-- a fully quantum mechanical treatment must take these into account \cite{RefWorks:doc:5a4d578de4b0c2a8b20ccff5,RefWorks:doc:5a15b8c4e4b0c5584aef9f41}. The series of diffraction integrals used to compute the CRB for the proposed interferometer are described in detail in \cite{RefWorks:doc:59d6a369e4b04214d8e4ea74}, with more specifics of the setup depicted in Fig. S4. The parameters $r_{\circ}$ and $M$ were chosen by computing $\sigma_z^{\text{(CRB)}}$ for a range of values (Fig. S5 of \cite{RefWorks:doc:59d6a369e4b04214d8e4ea74}). Applying a small phase correction at the Fourier plane before the annular mirror compensates for defocus accrued downstream. 

As seen in Fig. \ref{fig_sigma_single_obj}(b), this interferometer gives $\sigma_z^{\text{(CRB)}} \approx 1.03\times\sigma_z^{\text{(QCRB)}}$ near $z = 0$. The prefactor can be made closer to unity by incorporating additional beam splitter stages to make use of the essentially unused inner ring of the outer arm. We note that a relative deterioration in lateral precision accompanies the improvement in depth precision for this particular arrangement [Fig. \ref{fig_sigma_single_obj}(a)]. 

Since the signal is recorded in a conjugate Fourier plane and is not shift-invariant, the radial shear interferometer is not a viable configuration for wide-field imaging and is instead more compatible with confocal scanning or feedback-based particle tracking. For instance, it may be well-suited as an add-on to the MINFLUX microscope \cite{RefWorks:doc:59f64286e4b028f0a93ad056}, for which lateral position is determined solely by illumination modulation and so the detection optics can be reserved for depth estimation. A perhaps more experimentally attractive variation of the radial shear interferometer in which the signal is integrated onto three point detectors rather than two cameras is analyzed in \cite{RefWorks:doc:59d6a369e4b04214d8e4ea74} and gives $\sigma_z^{\text{(CRB)}} \approx 1.05\times\sigma_z^{\text{(QCRB)}}$ near $z = 0$.  Practical considerations aside, it is worthwhile to devise here a measurement scheme that approaches the ultimate bound.

\begin{figure}
\includegraphics[width=8.6cm]{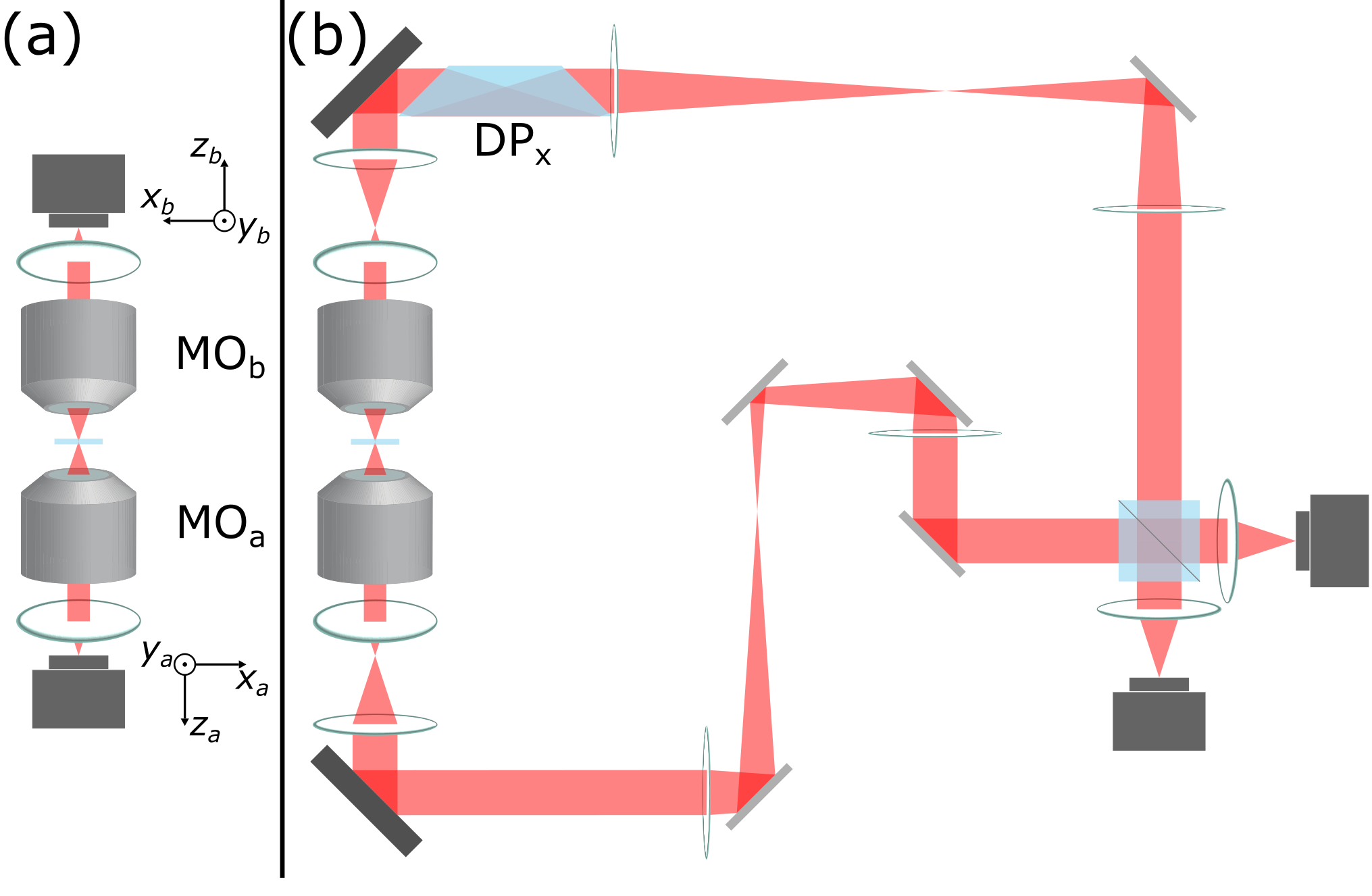}
\caption{Dual-objective microscope schematics. (a) Simplest dual-objective scheme in which the signal collected by microscope objectives $a$ and $b$ (MO\textsubscript{a}, MO\textsubscript{b}) is detected on two cameras without recombination. (b) Interferometric detection. Optimal lateral localization requires an additional reflection in one arm, as enforced, e.g., by an $x$-oriented Dove prism (DP\textsubscript{x}).}
\label{fig_dual_obj_sketches}
\end{figure}

Advanced fluorescence microscopy implementations sometimes make use of two opposed objectives (Fig. \ref{fig_dual_obj_sketches}) \cite{RefWorks:doc:59db9677e4b061a5d61518a3,RefWorks:doc:59db975ce4b00f3d39180384,RefWorks:doc:59db977ae4b0573d0abc0767,RefWorks:doc:59db9698e4b02fc0bca64201}. We also consider the quantum bounds for localization using this geometry, for which the state to be plugged into Eqs. (\ref{eq_QFI_def}) and (\ref{eq_SLD_def}) is given by $\rho(\mathbf{x}) = \ket{\psi(\mathbf{x})}\bra{\psi(\mathbf{x})}$ now with:
\begin{align}\label{eq_dual_obj_psi}
\ket{\psi(\mathbf{x})} = \frac{1}{\sqrt{2}}\iint \mathrm{d}A_F^{(a)} \psi\left(x_F^{(a)},y_F^{(a)};[x,y,z]^{\text{T}}\right) &\Ket{x_F^{(a)},y_F^{(a)}} \nonumber \\ + \frac{1}{\sqrt{2}}\iint \mathrm{d}A_F^{(b)} \psi\left(x_F^{(b)},y_F^{(b)};[-x,y,-z]^{\text{T}}\right) &\Ket{x_F^{(b)},y_F^{(b)}},
\end{align}
where superscript $(a)$ and $(b)$ refer to the coordinates at the back apertures of objectives $a$ and $b$ (Fig. \ref{fig_dual_obj_sketches}). 

\begin{figure}
\includegraphics[width=8.6cm]{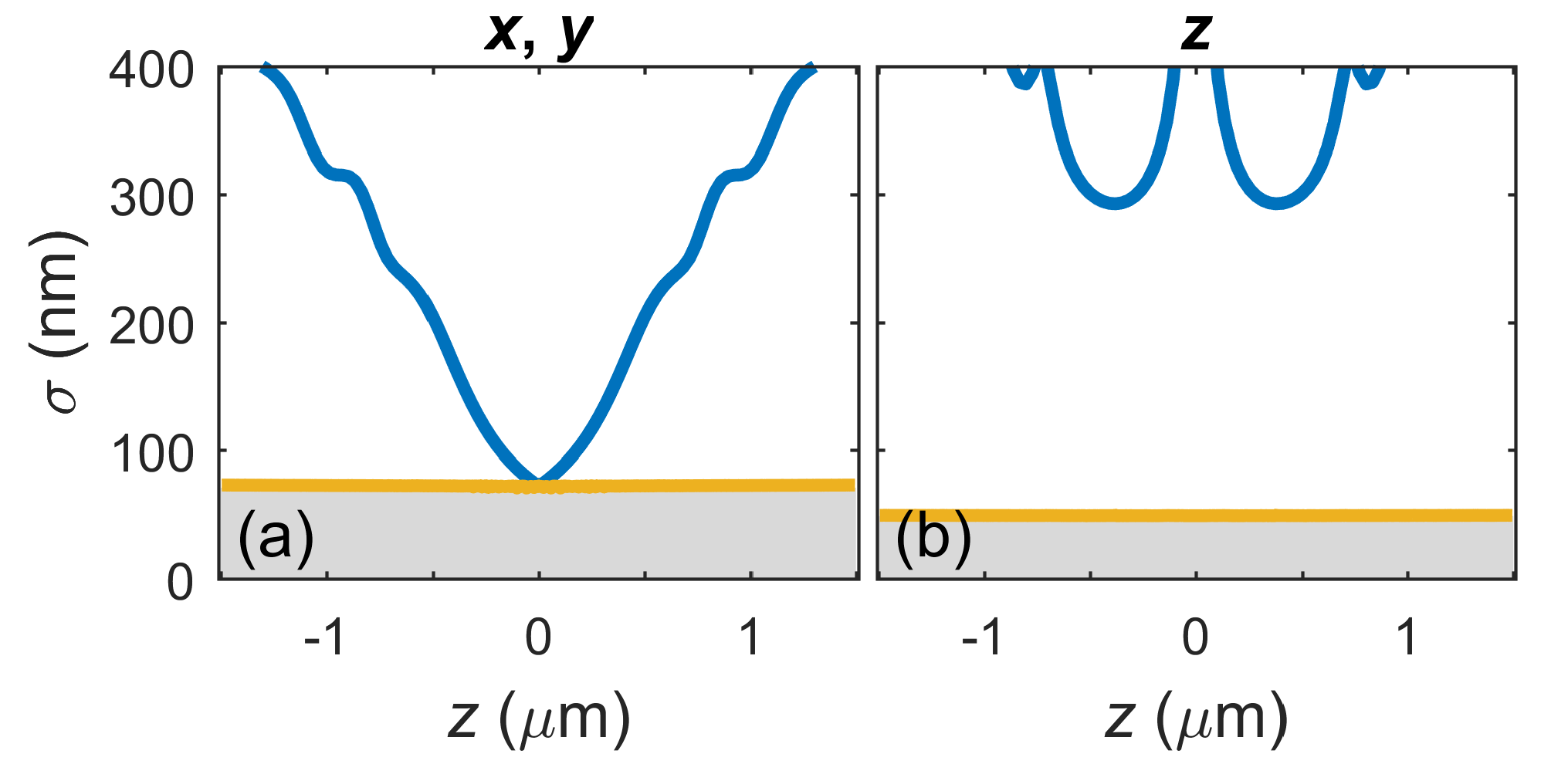}
\caption{Photon-normalized QCRBs and measurement CRBs for dual-objective detection as a function of source distance from focus. For $N$ detected signal photons divide vertical axis by $\sqrt{N}$. (a) Lateral localization bounds. The gray shaded region is bounded above by $\sigma = \sigma^{\text{(QCRB)}}_x = \sigma^{\text{(QCRB)}}_y$. Blue curve shows lateral CRB for non-interferometric detection and gold line corresponds to interferometric detection. The interferometric scheme globally achieves the QCRB. (b) Depth localization bounds. Color code corresponds to that in (a). Again the interferometric scheme obtains the bound for all $z$.}
\label{fig_sigma_dual_obj}
\end{figure}

The results are \cite{RefWorks:doc:59d6a369e4b04214d8e4ea74}:
\begin{subequations}
	\begin{align}
    	\sigma_x^{\text{(QCRB)}} &= \sigma_y^{\text{(QCRB)}} = C_{xy}/2, \\ \sigma_z^{\text{(QCRB)}} &= C_z/2,
    \end{align}
\end{subequations}
where $C_{xy}$ and $C_z$ are defined as before. Dual-objective QCRBs are depicted in Fig. \ref{fig_sigma_dual_obj}. In a real microscopy experiment the use of two objectives would double the rate of photon detections, but our normalized expressions scale this effect away. Thus, simply detecting with two cameras without further processing [Fig. \ref{fig_dual_obj_sketches}(a)] leads to the same CRBs as for the standard single-objective microscope (blue curves in Fig. \ref{fig_sigma_dual_obj}). A more sophisticated approach is to combine the signal due to objectives $a$ and $b$ interferometrically [Fig. \ref{fig_dual_obj_sketches}(b)], as in interferometric photo-activation localization microscopy (iPALM) \cite{RefWorks:doc:59db9698e4b02fc0bca64201}. Interferometric localization microscopy is known to produce superior depth localization precision relative to other common techniques \cite{RefWorks:doc:59db977ae4b0573d0abc0767}. Interestingly we find that in the considered limit of negligible background light, this configuration globally achieves the quantum bound in all three dimensions. This means that no additional optical elements incorporated into the setup in Fig. \ref{fig_dual_obj_sketches}(b) can lead to improved localization precision bounds. This is at odds with the naive notion that perhaps one can improve the depth localization precision bound by combining interferometric and PSF-engineering techniques. 

In conclusion, by deriving the QCRB for depth localization in a form relevant to advanced single-molecule microscopy techniques, we gained insight into the limits of commonly-used PSF-engineering approaches, subject to semiclassical photodetection in the limit of Poisson counting statistics. We showed that existing techniques fall short of the quantum bound for $z$ localization, and proposed a novel interferometer that can locally attain the bound. For dual-objective collection, we showed that an established interferometric detection method globally saturates the bounds for all three dimensions simultaneously. Finite background can be introduced by considering the appropriate mixed photon states, which we reserve for a future study. Our results are relevant for ongoing work on the 3D localization of sources of more complicated photon states, including both intense thermal states and distinctly nonclassical states.

Here, we have built upon previous work demonstrating the utility of quantum statistical approaches even for semiclassical microscopy of weak sources \cite{RefWorks:doc:59d68d25e4b04e4ae184ba79,RefWorks:doc:59d6a59ce4b0392c41dee5b0}. Future work in which the microscope's response function is engineered to increase information about source position (or any other estimandum, e.g., molecular orientation \cite{RefWorks:doc:58d6edb6e4b0b1f71a1fc890}) should be carried out with reference to the measurement-independent bounds.

\begin{acknowledgments}

This material is based upon work supported by, or in part by, the United States Army Research Laboratory and the United States Army Research Office under Grant No. W911NF1510548; as well as the Air Force Office of Scientific Research Grant. No. FA9550-17-1-0371. We thank Dikla Oren for helpful discussions.

\end{acknowledgments}

\bibliography{export_14_}

\begin{thebibliography}{72}%
\makeatletter
\providecommand \@ifxundefined [1]{%
 \@ifx{#1\undefined}
}%
\providecommand \@ifnum [1]{%
 \ifnum #1\expandafter \@firstoftwo
 \else \expandafter \@secondoftwo
 \fi
}%
\providecommand \@ifx [1]{%
 \ifx #1\expandafter \@firstoftwo
 \else \expandafter \@secondoftwo
 \fi
}%
\providecommand \natexlab [1]{#1}%
\providecommand \enquote  [1]{``#1''}%
\providecommand \bibnamefont  [1]{#1}%
\providecommand \bibfnamefont [1]{#1}%
\providecommand \citenamefont [1]{#1}%
\providecommand \href@noop [0]{\@secondoftwo}%
\providecommand \href [0]{\begingroup \@sanitize@url \@href}%
\providecommand \@href[1]{\@@startlink{#1}\@@href}%
\providecommand \@@href[1]{\endgroup#1\@@endlink}%
\providecommand \@sanitize@url [0]{\catcode `\\12\catcode `\$12\catcode
  `\&12\catcode `\#12\catcode `\^12\catcode `\_12\catcode `\%12\relax}%
\providecommand \@@startlink[1]{}%
\providecommand \@@endlink[0]{}%
\providecommand \url  [0]{\begingroup\@sanitize@url \@url }%
\providecommand \@url [1]{\endgroup\@href {#1}{\urlprefix }}%
\providecommand \urlprefix  [0]{URL }%
\providecommand \Eprint [0]{\href }%
\providecommand \doibase [0]{http://dx.doi.org/}%
\providecommand \selectlanguage [0]{\@gobble}%
\providecommand \bibinfo  [0]{\@secondoftwo}%
\providecommand \bibfield  [0]{\@secondoftwo}%
\providecommand \translation [1]{[#1]}%
\providecommand \BibitemOpen [0]{}%
\providecommand \bibitemStop [0]{}%
\providecommand \bibitemNoStop [0]{.\EOS\space}%
\providecommand \EOS [0]{\spacefactor3000\relax}%
\providecommand \BibitemShut  [1]{\csname bibitem#1\endcsname}%
\let\auto@bib@innerbib\@empty
\bibitem [{\citenamefont {Maurer}\ \emph {et~al.}(2010)\citenamefont {Maurer},
  \citenamefont {Maze}, \citenamefont {Stanwix}, \citenamefont {Jiang},
  \citenamefont {Gorshkov}, \citenamefont {Zibrov}, \citenamefont {Harke},
  \citenamefont {Hodges}, \citenamefont {Zibrov},\ and\ \citenamefont
  {Yacoby}}]{RefWorks:doc:5a96d632e4b04ec5e747c2a2}%
  \BibitemOpen
  \bibfield  {author} {\bibinfo {author} {\bibfnamefont {P.~C.}\ \bibnamefont
  {Maurer}}, \bibinfo {author} {\bibfnamefont {J.~R.}\ \bibnamefont {Maze}},
  \bibinfo {author} {\bibfnamefont {P.~L.}\ \bibnamefont {Stanwix}}, \bibinfo
  {author} {\bibfnamefont {L.}~\bibnamefont {Jiang}}, \bibinfo {author}
  {\bibfnamefont {A.~V.}\ \bibnamefont {Gorshkov}}, \bibinfo {author}
  {\bibfnamefont {A.~A.}\ \bibnamefont {Zibrov}}, \bibinfo {author}
  {\bibfnamefont {B.}~\bibnamefont {Harke}}, \bibinfo {author} {\bibfnamefont
  {J.~S.}\ \bibnamefont {Hodges}}, \bibinfo {author} {\bibfnamefont {A.~S.}\
  \bibnamefont {Zibrov}}, \ and\ \bibinfo {author} {\bibfnamefont
  {A.}~\bibnamefont {Yacoby}},\ }\href@noop {} {\bibfield  {journal} {\bibinfo
  {journal} {Nature Physics}\ }\textbf {\bibinfo {volume} {6}},\ \bibinfo
  {pages} {912} (\bibinfo {year} {2010})}\BibitemShut {NoStop}%
\bibitem [{\citenamefont {Arai}\ \emph {et~al.}(2015)\citenamefont {Arai},
  \citenamefont {Belthangady}, \citenamefont {Zhang}, \citenamefont {Bar-Gill},
  \citenamefont {DeVience}, \citenamefont {Cappellaro}, \citenamefont
  {Yacoby},\ and\ \citenamefont
  {Walsworth}}]{RefWorks:doc:5a96d66ae4b027b722a9bba8}%
  \BibitemOpen
  \bibfield  {author} {\bibinfo {author} {\bibfnamefont {K.}~\bibnamefont
  {Arai}}, \bibinfo {author} {\bibfnamefont {C.}~\bibnamefont {Belthangady}},
  \bibinfo {author} {\bibfnamefont {H.}~\bibnamefont {Zhang}}, \bibinfo
  {author} {\bibfnamefont {N.}~\bibnamefont {Bar-Gill}}, \bibinfo {author}
  {\bibfnamefont {S.~J.}\ \bibnamefont {DeVience}}, \bibinfo {author}
  {\bibfnamefont {P.}~\bibnamefont {Cappellaro}}, \bibinfo {author}
  {\bibfnamefont {A.}~\bibnamefont {Yacoby}}, \ and\ \bibinfo {author}
  {\bibfnamefont {R.~L.}\ \bibnamefont {Walsworth}},\ }\href@noop {} {\bibfield
   {journal} {\bibinfo  {journal} {Nature Nanotechnology}\ }\textbf {\bibinfo
  {volume} {10}},\ \bibinfo {pages} {859} (\bibinfo {year} {2015})}\BibitemShut
  {NoStop}%
\bibitem [{\citenamefont {Jaskula}\ \emph {et~al.}(2017)\citenamefont
  {Jaskula}, \citenamefont {Bauch}, \citenamefont {Arroyo-Camejo},
  \citenamefont {Lukin}, \citenamefont {Hell}, \citenamefont {Trifonov},\ and\
  \citenamefont {Walsworth}}]{RefWorks:doc:5a96d6a7e4b0642c4190f839}%
  \BibitemOpen
  \bibfield  {author} {\bibinfo {author} {\bibfnamefont {J.-C.}\ \bibnamefont
  {Jaskula}}, \bibinfo {author} {\bibfnamefont {E.}~\bibnamefont {Bauch}},
  \bibinfo {author} {\bibfnamefont {S.}~\bibnamefont {Arroyo-Camejo}}, \bibinfo
  {author} {\bibfnamefont {M.~D.}\ \bibnamefont {Lukin}}, \bibinfo {author}
  {\bibfnamefont {S.~W.}\ \bibnamefont {Hell}}, \bibinfo {author}
  {\bibfnamefont {A.~S.}\ \bibnamefont {Trifonov}}, \ and\ \bibinfo {author}
  {\bibfnamefont {R.~L.}\ \bibnamefont {Walsworth}},\ }\href@noop {} {\bibfield
   {journal} {\bibinfo  {journal} {Optics Express}\ }\textbf {\bibinfo {volume}
  {25}},\ \bibinfo {pages} {11048} (\bibinfo {year} {2017})}\BibitemShut
  {NoStop}%
\bibitem [{\citenamefont {Zhang}\ \emph {et~al.}(2017)\citenamefont {Zhang},
  \citenamefont {Arai}, \citenamefont {Belthangady}, \citenamefont {Jaskula},\
  and\ \citenamefont {Walsworth}}]{RefWorks:doc:5a96d6f2e4b0642c4190f85e}%
  \BibitemOpen
  \bibfield  {author} {\bibinfo {author} {\bibfnamefont {H.}~\bibnamefont
  {Zhang}}, \bibinfo {author} {\bibfnamefont {K.}~\bibnamefont {Arai}},
  \bibinfo {author} {\bibfnamefont {C.}~\bibnamefont {Belthangady}}, \bibinfo
  {author} {\bibfnamefont {J.-C.}\ \bibnamefont {Jaskula}}, \ and\ \bibinfo
  {author} {\bibfnamefont {R.~L.}\ \bibnamefont {Walsworth}},\ }\href@noop {}
  {\bibfield  {journal} {\bibinfo  {journal} {NPJ Quantum Information}\
  }\textbf {\bibinfo {volume} {3}},\ \bibinfo {pages} {31} (\bibinfo {year}
  {2017})}\BibitemShut {NoStop}%
\bibitem [{\citenamefont {Deschout}\ \emph {et~al.}(2014)\citenamefont
  {Deschout}, \citenamefont {Zanacchi}, \citenamefont {Mlodzianoski},
  \citenamefont {Diaspro}, \citenamefont {Bewersdorf}, \citenamefont {Hess},\
  and\ \citenamefont {Braeckmans}}]{RefWorks:doc:58d6ed0ce4b05fe42c93d941}%
  \BibitemOpen
  \bibfield  {author} {\bibinfo {author} {\bibfnamefont {H.}~\bibnamefont
  {Deschout}}, \bibinfo {author} {\bibfnamefont {F.~C.}\ \bibnamefont
  {Zanacchi}}, \bibinfo {author} {\bibfnamefont {M.}~\bibnamefont
  {Mlodzianoski}}, \bibinfo {author} {\bibfnamefont {A.}~\bibnamefont
  {Diaspro}}, \bibinfo {author} {\bibfnamefont {J.}~\bibnamefont {Bewersdorf}},
  \bibinfo {author} {\bibfnamefont {S.~T.}\ \bibnamefont {Hess}}, \ and\
  \bibinfo {author} {\bibfnamefont {K.}~\bibnamefont {Braeckmans}},\
  }\href@noop {} {\bibfield  {journal} {\bibinfo  {journal} {Nature Methods}\
  }\textbf {\bibinfo {volume} {11}},\ \bibinfo {pages} {253} (\bibinfo {year}
  {2014})}\BibitemShut {NoStop}%
\bibitem [{\citenamefont {Shen}\ \emph {et~al.}(2017)\citenamefont {Shen},
  \citenamefont {Tauzin}, \citenamefont {Baiyasi}, \citenamefont {Wang},
  \citenamefont {Moringo}, \citenamefont {Shuang},\ and\ \citenamefont
  {Landes}}]{RefWorks:doc:59d57285e4b0954cd3d740dd}%
  \BibitemOpen
  \bibfield  {author} {\bibinfo {author} {\bibfnamefont {H.}~\bibnamefont
  {Shen}}, \bibinfo {author} {\bibfnamefont {L.~J.}\ \bibnamefont {Tauzin}},
  \bibinfo {author} {\bibfnamefont {R.}~\bibnamefont {Baiyasi}}, \bibinfo
  {author} {\bibfnamefont {W.}~\bibnamefont {Wang}}, \bibinfo {author}
  {\bibfnamefont {N.}~\bibnamefont {Moringo}}, \bibinfo {author} {\bibfnamefont
  {B.}~\bibnamefont {Shuang}}, \ and\ \bibinfo {author} {\bibfnamefont {C.~F.}\
  \bibnamefont {Landes}},\ }\href@noop {} {\bibfield  {journal} {\bibinfo
  {journal} {Chemical Reviews}\ } (\bibinfo {year} {2017})}\BibitemShut
  {NoStop}%
\bibitem [{\citenamefont {von Diezmann}\ \emph {et~al.}(2017)\citenamefont {von
  Diezmann}, \citenamefont {Shechtman},\ and\ \citenamefont
  {Moerner}}]{RefWorks:doc:59d57455e4b07a74c3b17159}%
  \BibitemOpen
  \bibfield  {author} {\bibinfo {author} {\bibfnamefont {A.}~\bibnamefont {von
  Diezmann}}, \bibinfo {author} {\bibfnamefont {Y.}~\bibnamefont {Shechtman}},
  \ and\ \bibinfo {author} {\bibfnamefont {W.~E.}\ \bibnamefont {Moerner}},\
  }\href@noop {} {\bibfield  {journal} {\bibinfo  {journal} {Chemical Reviews}\
  } (\bibinfo {year} {2017})}\BibitemShut {NoStop}%
\bibitem [{\citenamefont {Prabhat}\ \emph {et~al.}(2004)\citenamefont
  {Prabhat}, \citenamefont {Ram}, \citenamefont {Ward},\ and\ \citenamefont
  {Ober}}]{RefWorks:doc:5a09c164e4b0e70d304d6a26}%
  \BibitemOpen
  \bibfield  {author} {\bibinfo {author} {\bibfnamefont {P.}~\bibnamefont
  {Prabhat}}, \bibinfo {author} {\bibfnamefont {S.}~\bibnamefont {Ram}},
  \bibinfo {author} {\bibfnamefont {E.~S.}\ \bibnamefont {Ward}}, \ and\
  \bibinfo {author} {\bibfnamefont {R.~J.}\ \bibnamefont {Ober}},\ }\href@noop
  {} {\bibfield  {journal} {\bibinfo  {journal} {IEEE Transactions on
  Nanobioscience}\ }\textbf {\bibinfo {volume} {3}},\ \bibinfo {pages} {237}
  (\bibinfo {year} {2004})}\BibitemShut {NoStop}%
\bibitem [{\citenamefont {Huang}\ \emph {et~al.}(2008)\citenamefont {Huang},
  \citenamefont {Wang}, \citenamefont {Bates},\ and\ \citenamefont
  {Zhuang}}]{RefWorks:doc:586d17b2e4b00198145b4b75}%
  \BibitemOpen
  \bibfield  {author} {\bibinfo {author} {\bibfnamefont {B.}~\bibnamefont
  {Huang}}, \bibinfo {author} {\bibfnamefont {W.}~\bibnamefont {Wang}},
  \bibinfo {author} {\bibfnamefont {M.}~\bibnamefont {Bates}}, \ and\ \bibinfo
  {author} {\bibfnamefont {X.}~\bibnamefont {Zhuang}},\ }\href@noop {}
  {\bibfield  {journal} {\bibinfo  {journal} {Science}\ }\textbf {\bibinfo
  {volume} {319}},\ \bibinfo {pages} {810} (\bibinfo {year}
  {2008})}\BibitemShut {NoStop}%
\bibitem [{\citenamefont {Piestun}\ \emph {et~al.}(2000)\citenamefont
  {Piestun}, \citenamefont {Schechner},\ and\ \citenamefont
  {Shamir}}]{RefWorks:doc:59d68600e4b0573d0abb7158}%
  \BibitemOpen
  \bibfield  {author} {\bibinfo {author} {\bibfnamefont {R.}~\bibnamefont
  {Piestun}}, \bibinfo {author} {\bibfnamefont {Y.~Y.}\ \bibnamefont
  {Schechner}}, \ and\ \bibinfo {author} {\bibfnamefont {J.}~\bibnamefont
  {Shamir}},\ }\href@noop {} {\bibfield  {journal} {\bibinfo  {journal} {JOSA
  A}\ }\textbf {\bibinfo {volume} {17}},\ \bibinfo {pages} {294} (\bibinfo
  {year} {2000})}\BibitemShut {NoStop}%
\bibitem [{\citenamefont {Pavani}\ \emph {et~al.}(2009)\citenamefont {Pavani},
  \citenamefont {Thompson}, \citenamefont {Biteen}, \citenamefont {Lord},
  \citenamefont {Liu}, \citenamefont {Twieg}, \citenamefont {Piestun},\ and\
  \citenamefont {Moerner}}]{RefWorks:doc:586d17c9e4b0147c58343dfc}%
  \BibitemOpen
  \bibfield  {author} {\bibinfo {author} {\bibfnamefont {S.~R.~P.}\
  \bibnamefont {Pavani}}, \bibinfo {author} {\bibfnamefont {M.~A.}\
  \bibnamefont {Thompson}}, \bibinfo {author} {\bibfnamefont {J.~S.}\
  \bibnamefont {Biteen}}, \bibinfo {author} {\bibfnamefont {S.~J.}\
  \bibnamefont {Lord}}, \bibinfo {author} {\bibfnamefont {N.}~\bibnamefont
  {Liu}}, \bibinfo {author} {\bibfnamefont {R.~J.}\ \bibnamefont {Twieg}},
  \bibinfo {author} {\bibfnamefont {R.}~\bibnamefont {Piestun}}, \ and\
  \bibinfo {author} {\bibfnamefont {W.~E.}\ \bibnamefont {Moerner}},\
  }\href@noop {} {\bibfield  {journal} {\bibinfo  {journal} {Proceedings of the
  National Academy of Sciences}\ }\textbf {\bibinfo {volume} {106}},\ \bibinfo
  {pages} {2995} (\bibinfo {year} {2009})}\BibitemShut {NoStop}%
\bibitem [{\citenamefont {Jia}\ \emph {et~al.}(2014)\citenamefont {Jia},
  \citenamefont {Vaughan},\ and\ \citenamefont
  {Zhuang}}]{RefWorks:doc:59d68668e4b061a5d6147ad6}%
  \BibitemOpen
  \bibfield  {author} {\bibinfo {author} {\bibfnamefont {S.}~\bibnamefont
  {Jia}}, \bibinfo {author} {\bibfnamefont {J.~C.}\ \bibnamefont {Vaughan}}, \
  and\ \bibinfo {author} {\bibfnamefont {X.}~\bibnamefont {Zhuang}},\
  }\href@noop {} {\bibfield  {journal} {\bibinfo  {journal} {Nature Photonics}\
  }\textbf {\bibinfo {volume} {8}},\ \bibinfo {pages} {302} (\bibinfo {year}
  {2014})}\BibitemShut {NoStop}%
\bibitem [{\citenamefont {Lew}\ \emph {et~al.}(2011)\citenamefont {Lew},
  \citenamefont {Lee}, \citenamefont {Badieirostami},\ and\ \citenamefont
  {Moerner}}]{RefWorks:doc:59d685d3e4b061a5d6147ac2}%
  \BibitemOpen
  \bibfield  {author} {\bibinfo {author} {\bibfnamefont {M.~D.}\ \bibnamefont
  {Lew}}, \bibinfo {author} {\bibfnamefont {S.~F.}\ \bibnamefont {Lee}},
  \bibinfo {author} {\bibfnamefont {M.}~\bibnamefont {Badieirostami}}, \ and\
  \bibinfo {author} {\bibfnamefont {W.~E.}\ \bibnamefont {Moerner}},\
  }\href@noop {} {\bibfield  {journal} {\bibinfo  {journal} {Optics Letters}\
  }\textbf {\bibinfo {volume} {36}},\ \bibinfo {pages} {202} (\bibinfo {year}
  {2011})}\BibitemShut {NoStop}%
\bibitem [{\citenamefont {Baddeley}\ \emph {et~al.}(2011)\citenamefont
  {Baddeley}, \citenamefont {Cannell},\ and\ \citenamefont
  {Soeller}}]{RefWorks:doc:59d68638e4b00aed3eaf3cf5}%
  \BibitemOpen
  \bibfield  {author} {\bibinfo {author} {\bibfnamefont {D.}~\bibnamefont
  {Baddeley}}, \bibinfo {author} {\bibfnamefont {M.~B.}\ \bibnamefont
  {Cannell}}, \ and\ \bibinfo {author} {\bibfnamefont {C.}~\bibnamefont
  {Soeller}},\ }\href@noop {} {\bibfield  {journal} {\bibinfo  {journal} {Nano
  Research}\ }\textbf {\bibinfo {volume} {4}},\ \bibinfo {pages} {589}
  (\bibinfo {year} {2011})}\BibitemShut {NoStop}%
\bibitem [{\citenamefont {Juette}\ \emph {et~al.}(2008)\citenamefont {Juette},
  \citenamefont {Gould}, \citenamefont {Lessard}, \citenamefont {Mlodzianoski},
  \citenamefont {Nagpure}, \citenamefont {Bennett}, \citenamefont {Hess},\ and\
  \citenamefont {Bewersdorf}}]{RefWorks:doc:59d6885ae4b04214d8e4e5e9}%
  \BibitemOpen
  \bibfield  {author} {\bibinfo {author} {\bibfnamefont {M.~F.}\ \bibnamefont
  {Juette}}, \bibinfo {author} {\bibfnamefont {T.~J.}\ \bibnamefont {Gould}},
  \bibinfo {author} {\bibfnamefont {M.~D.}\ \bibnamefont {Lessard}}, \bibinfo
  {author} {\bibfnamefont {M.~J.}\ \bibnamefont {Mlodzianoski}}, \bibinfo
  {author} {\bibfnamefont {B.~S.}\ \bibnamefont {Nagpure}}, \bibinfo {author}
  {\bibfnamefont {B.~T.}\ \bibnamefont {Bennett}}, \bibinfo {author}
  {\bibfnamefont {S.~T.}\ \bibnamefont {Hess}}, \ and\ \bibinfo {author}
  {\bibfnamefont {J.}~\bibnamefont {Bewersdorf}},\ }\href@noop {} {\bibfield
  {journal} {\bibinfo  {journal} {Nature Methods}\ }\textbf {\bibinfo {volume}
  {5}},\ \bibinfo {pages} {527} (\bibinfo {year} {2008})}\BibitemShut {NoStop}%
\bibitem [{\citenamefont {Abrahamsson}\ \emph {et~al.}(2013)\citenamefont
  {Abrahamsson}, \citenamefont {Chen}, \citenamefont {Hajj}, \citenamefont
  {Stallinga}, \citenamefont {Katsov}, \citenamefont {Wisniewski},
  \citenamefont {Mizuguchi}, \citenamefont {Soule}, \citenamefont {Mueller},\
  and\ \citenamefont {Darzacq}}]{RefWorks:doc:59d6888de4b04214d8e4e5ee}%
  \BibitemOpen
  \bibfield  {author} {\bibinfo {author} {\bibfnamefont {S.}~\bibnamefont
  {Abrahamsson}}, \bibinfo {author} {\bibfnamefont {J.}~\bibnamefont {Chen}},
  \bibinfo {author} {\bibfnamefont {B.}~\bibnamefont {Hajj}}, \bibinfo {author}
  {\bibfnamefont {S.}~\bibnamefont {Stallinga}}, \bibinfo {author}
  {\bibfnamefont {A.~Y.}\ \bibnamefont {Katsov}}, \bibinfo {author}
  {\bibfnamefont {J.}~\bibnamefont {Wisniewski}}, \bibinfo {author}
  {\bibfnamefont {G.}~\bibnamefont {Mizuguchi}}, \bibinfo {author}
  {\bibfnamefont {P.}~\bibnamefont {Soule}}, \bibinfo {author} {\bibfnamefont
  {F.}~\bibnamefont {Mueller}}, \ and\ \bibinfo {author} {\bibfnamefont
  {C.~D.}\ \bibnamefont {Darzacq}},\ }\href@noop {} {\bibfield  {journal}
  {\bibinfo  {journal} {Nature Methods}\ }\textbf {\bibinfo {volume} {10}},\
  \bibinfo {pages} {60} (\bibinfo {year} {2013})}\BibitemShut {NoStop}%
\bibitem [{\citenamefont {Hell}\ \emph {et~al.}(1994)\citenamefont {Hell},
  \citenamefont {Lindek}, \citenamefont {Cremer},\ and\ \citenamefont
  {Stelzer}}]{RefWorks:doc:59db9677e4b061a5d61518a3}%
  \BibitemOpen
  \bibfield  {author} {\bibinfo {author} {\bibfnamefont {S.~W.}\ \bibnamefont
  {Hell}}, \bibinfo {author} {\bibfnamefont {S.}~\bibnamefont {Lindek}},
  \bibinfo {author} {\bibfnamefont {C.}~\bibnamefont {Cremer}}, \ and\ \bibinfo
  {author} {\bibfnamefont {E.~H.}\ \bibnamefont {Stelzer}},\ }\href@noop {}
  {\bibfield  {journal} {\bibinfo  {journal} {Optics Letters}\ }\textbf
  {\bibinfo {volume} {19}},\ \bibinfo {pages} {222} (\bibinfo {year}
  {1994})}\BibitemShut {NoStop}%
\bibitem [{\citenamefont {Gustafsson}\ \emph {et~al.}(1999)\citenamefont
  {Gustafsson}, \citenamefont {Agard},\ and\ \citenamefont
  {Sedat}}]{RefWorks:doc:59db975ce4b00f3d39180384}%
  \BibitemOpen
  \bibfield  {author} {\bibinfo {author} {\bibfnamefont {M.~G.}\ \bibnamefont
  {Gustafsson}}, \bibinfo {author} {\bibfnamefont {D.~A.}\ \bibnamefont
  {Agard}}, \ and\ \bibinfo {author} {\bibfnamefont {J.~W.}\ \bibnamefont
  {Sedat}},\ }\href@noop {} {\bibfield  {journal} {\bibinfo  {journal} {Journal
  of Microscopy}\ }\textbf {\bibinfo {volume} {195}},\ \bibinfo {pages} {10}
  (\bibinfo {year} {1999})}\BibitemShut {NoStop}%
\bibitem [{\citenamefont {v.~Middendorff}\ \emph {et~al.}(2008)\citenamefont
  {v.~Middendorff}, \citenamefont {Egner}, \citenamefont {Geisler},
  \citenamefont {Hell},\ and\ \citenamefont
  {Schnle}}]{RefWorks:doc:59db977ae4b0573d0abc0767}%
  \BibitemOpen
  \bibfield  {author} {\bibinfo {author} {\bibfnamefont {C.}~\bibnamefont
  {v.~Middendorff}}, \bibinfo {author} {\bibfnamefont {A.}~\bibnamefont
  {Egner}}, \bibinfo {author} {\bibfnamefont {C.}~\bibnamefont {Geisler}},
  \bibinfo {author} {\bibfnamefont {S.}~\bibnamefont {Hell}}, \ and\ \bibinfo
  {author} {\bibfnamefont {A.}~\bibnamefont {Schnle}},\ }\href@noop {}
  {\bibfield  {journal} {\bibinfo  {journal} {Optics Express}\ }\textbf
  {\bibinfo {volume} {16}},\ \bibinfo {pages} {20774} (\bibinfo {year}
  {2008})}\BibitemShut {NoStop}%
\bibitem [{\citenamefont {Shtengel}\ \emph {et~al.}(2009)\citenamefont
  {Shtengel}, \citenamefont {Galbraith}, \citenamefont {Galbraith},
  \citenamefont {Lippincott-Schwartz}, \citenamefont {Gillette}, \citenamefont
  {Manley}, \citenamefont {Sougrat}, \citenamefont {Waterman}, \citenamefont
  {Kanchanawong},\ and\ \citenamefont
  {Davidson}}]{RefWorks:doc:59db9698e4b02fc0bca64201}%
  \BibitemOpen
  \bibfield  {author} {\bibinfo {author} {\bibfnamefont {G.}~\bibnamefont
  {Shtengel}}, \bibinfo {author} {\bibfnamefont {J.~A.}\ \bibnamefont
  {Galbraith}}, \bibinfo {author} {\bibfnamefont {C.~G.}\ \bibnamefont
  {Galbraith}}, \bibinfo {author} {\bibfnamefont {J.}~\bibnamefont
  {Lippincott-Schwartz}}, \bibinfo {author} {\bibfnamefont {J.~M.}\
  \bibnamefont {Gillette}}, \bibinfo {author} {\bibfnamefont {S.}~\bibnamefont
  {Manley}}, \bibinfo {author} {\bibfnamefont {R.}~\bibnamefont {Sougrat}},
  \bibinfo {author} {\bibfnamefont {C.~M.}\ \bibnamefont {Waterman}}, \bibinfo
  {author} {\bibfnamefont {P.}~\bibnamefont {Kanchanawong}}, \ and\ \bibinfo
  {author} {\bibfnamefont {M.~W.}\ \bibnamefont {Davidson}},\ }\href@noop {}
  {\bibfield  {journal} {\bibinfo  {journal} {Proceedings of the National
  Academy of Sciences}\ }\textbf {\bibinfo {volume} {106}},\ \bibinfo {pages}
  {3125} (\bibinfo {year} {2009})}\BibitemShut {NoStop}%
\bibitem [{\citenamefont {Cover}\ and\ \citenamefont
  {Thomas}(2012)}]{RefWorks:doc:59d646f8e4b04e4ae184a962}%
  \BibitemOpen
  \bibfield  {author} {\bibinfo {author} {\bibfnamefont {T.~M.}\ \bibnamefont
  {Cover}}\ and\ \bibinfo {author} {\bibfnamefont {J.~A.}\ \bibnamefont
  {Thomas}},\ }\href@noop {} {\emph {\bibinfo {title} {Elements of Information
  Theory}}}\ (\bibinfo  {publisher} {John Wiley and Sons},\ \bibinfo {year}
  {2012})\BibitemShut {NoStop}%
\bibitem [{\citenamefont
  {Helstrom}(1976)}]{RefWorks:doc:59d568aee4b08398efcdb1f5}%
  \BibitemOpen
  \bibfield  {author} {\bibinfo {author} {\bibfnamefont {C.~W.}\ \bibnamefont
  {Helstrom}},\ }\href@noop {} {\emph {\bibinfo {title} {Quantum Detection and
  Estimation Theory}}}\ (\bibinfo  {publisher} {Academic Press},\ \bibinfo
  {year} {1976})\BibitemShut {NoStop}%
\bibitem [{\citenamefont {Mandel}\ and\ \citenamefont
  {Wolf}(1995)}]{RefWorks:doc:5a4d578de4b0c2a8b20ccff5}%
  \BibitemOpen
  \bibfield  {author} {\bibinfo {author} {\bibfnamefont {L.}~\bibnamefont
  {Mandel}}\ and\ \bibinfo {author} {\bibfnamefont {E.}~\bibnamefont {Wolf}},\
  }\href@noop {} {\emph {\bibinfo {title} {Optical Coherence and Quantum
  Optics}}}\ (\bibinfo  {publisher} {Cambridge University Press},\ \bibinfo
  {year} {1995})\BibitemShut {NoStop}%
\bibitem [{\citenamefont
  {Goodman}(2015)}]{RefWorks:doc:59d6b3a0e4b0573d0abb7a2c}%
  \BibitemOpen
  \bibfield  {author} {\bibinfo {author} {\bibfnamefont {J.~W.}\ \bibnamefont
  {Goodman}},\ }\href@noop {} {\emph {\bibinfo {title} {Statistical Optics}}}\
  (\bibinfo  {publisher} {John Wiley and Sons},\ \bibinfo {year}
  {2015})\BibitemShut {NoStop}%
\bibitem [{\citenamefont {Tsang}\ \emph
  {et~al.}(2016{\natexlab{a}})\citenamefont {Tsang}, \citenamefont {Nair},\
  and\ \citenamefont {Lu}}]{RefWorks:doc:59d68d25e4b04e4ae184ba79}%
  \BibitemOpen
  \bibfield  {author} {\bibinfo {author} {\bibfnamefont {M.}~\bibnamefont
  {Tsang}}, \bibinfo {author} {\bibfnamefont {R.}~\bibnamefont {Nair}}, \ and\
  \bibinfo {author} {\bibfnamefont {X.-M.}\ \bibnamefont {Lu}},\ }\href@noop {}
  {\bibfield  {journal} {\bibinfo  {journal} {Physical Review X}\ }\textbf
  {\bibinfo {volume} {6}},\ \bibinfo {pages} {031033} (\bibinfo {year}
  {2016}{\natexlab{a}})}\BibitemShut {NoStop}%
\bibitem [{\citenamefont {Tsang}\ \emph
  {et~al.}(2016{\natexlab{b}})\citenamefont {Tsang}, \citenamefont {Nair},\
  and\ \citenamefont {Lu}}]{RefWorks:doc:59d6a59ce4b0392c41dee5b0}%
  \BibitemOpen
  \bibfield  {author} {\bibinfo {author} {\bibfnamefont {M.}~\bibnamefont
  {Tsang}}, \bibinfo {author} {\bibfnamefont {R.}~\bibnamefont {Nair}}, \ and\
  \bibinfo {author} {\bibfnamefont {X.-M.}\ \bibnamefont {Lu}},\ }in\
  \href@noop {} {\emph {\bibinfo {booktitle} {Proc. SPIE}}},\ Vol.\ \bibinfo
  {volume} {10029}\ (\bibinfo {year} {2016})\ p.\ \bibinfo {pages}
  {1002903}\BibitemShut {NoStop}%
\bibitem [{\citenamefont
  {Tsang}(2018)}]{RefWorks:doc:5a96ba60e4b0952b36e61f25}%
  \BibitemOpen
  \bibfield  {author} {\bibinfo {author} {\bibfnamefont {M.}~\bibnamefont
  {Tsang}},\ }\href@noop {} {\bibfield  {journal} {\bibinfo  {journal}
  {Physical Review A}\ }\textbf {\bibinfo {volume} {97}},\ \bibinfo {pages}
  {023830} (\bibinfo {year} {2018})}\BibitemShut {NoStop}%
\bibitem [{\citenamefont {Ober}\ \emph {et~al.}(2004)\citenamefont {Ober},
  \citenamefont {Ram},\ and\ \citenamefont
  {Ward}}]{RefWorks:doc:59d64167e4b00aed3eaf28ca}%
  \BibitemOpen
  \bibfield  {author} {\bibinfo {author} {\bibfnamefont {R.~J.}\ \bibnamefont
  {Ober}}, \bibinfo {author} {\bibfnamefont {S.}~\bibnamefont {Ram}}, \ and\
  \bibinfo {author} {\bibfnamefont {E.~S.}\ \bibnamefont {Ward}},\ }\href@noop
  {} {\bibfield  {journal} {\bibinfo  {journal} {Biophysical Journal}\ }\textbf
  {\bibinfo {volume} {86}},\ \bibinfo {pages} {1185} (\bibinfo {year}
  {2004})}\BibitemShut {NoStop}%
\bibitem [{\citenamefont {Ram}\ \emph {et~al.}(2006)\citenamefont {Ram},
  \citenamefont {Ward},\ and\ \citenamefont
  {Ober}}]{RefWorks:doc:59d6414de4b061a5d61462a8}%
  \BibitemOpen
  \bibfield  {author} {\bibinfo {author} {\bibfnamefont {S.}~\bibnamefont
  {Ram}}, \bibinfo {author} {\bibfnamefont {E.~S.}\ \bibnamefont {Ward}}, \
  and\ \bibinfo {author} {\bibfnamefont {R.~J.}\ \bibnamefont {Ober}},\
  }\href@noop {} {\bibfield  {journal} {\bibinfo  {journal} {Proceedings of the
  National Academy of Sciences of the United States of America}\ }\textbf
  {\bibinfo {volume} {103}},\ \bibinfo {pages} {4457} (\bibinfo {year}
  {2006})}\BibitemShut {NoStop}%
\bibitem [{\citenamefont {Badieirostami}\ \emph {et~al.}(2010)\citenamefont
  {Badieirostami}, \citenamefont {Lew}, \citenamefont {Thompson},\ and\
  \citenamefont {Moerner}}]{RefWorks:doc:59d64194e4b061a5d61462b5}%
  \BibitemOpen
  \bibfield  {author} {\bibinfo {author} {\bibfnamefont {M.}~\bibnamefont
  {Badieirostami}}, \bibinfo {author} {\bibfnamefont {M.~D.}\ \bibnamefont
  {Lew}}, \bibinfo {author} {\bibfnamefont {M.~A.}\ \bibnamefont {Thompson}}, \
  and\ \bibinfo {author} {\bibfnamefont {W.~E.}\ \bibnamefont {Moerner}},\
  }\href@noop {} {\bibfield  {journal} {\bibinfo  {journal} {Applied Physics
  Letters}\ }\textbf {\bibinfo {volume} {97}},\ \bibinfo {pages} {161103}
  (\bibinfo {year} {2010})}\BibitemShut {NoStop}%
\bibitem [{\citenamefont {Shechtman}\ \emph {et~al.}(2014)\citenamefont
  {Shechtman}, \citenamefont {Sahl}, \citenamefont {Backer},\ and\
  \citenamefont {Moerner}}]{RefWorks:doc:586d17d9e4b00198145b4b7f}%
  \BibitemOpen
  \bibfield  {author} {\bibinfo {author} {\bibfnamefont {Y.}~\bibnamefont
  {Shechtman}}, \bibinfo {author} {\bibfnamefont {S.~J.}\ \bibnamefont {Sahl}},
  \bibinfo {author} {\bibfnamefont {A.~S.}\ \bibnamefont {Backer}}, \ and\
  \bibinfo {author} {\bibfnamefont {W.~E.}\ \bibnamefont {Moerner}},\
  }\href@noop {} {\bibfield  {journal} {\bibinfo  {journal} {Physical Review
  Letters}\ }\textbf {\bibinfo {volume} {113}},\ \bibinfo {pages} {133902}
  (\bibinfo {year} {2014})}\BibitemShut {NoStop}%
\bibitem [{\citenamefont {Shechtman}\ \emph {et~al.}(2015)\citenamefont
  {Shechtman}, \citenamefont {Weiss}, \citenamefont {Backer}, \citenamefont
  {Sahl},\ and\ \citenamefont
  {Moerner}}]{RefWorks:doc:586d17e0e4b00198145b4b94}%
  \BibitemOpen
  \bibfield  {author} {\bibinfo {author} {\bibfnamefont {Y.}~\bibnamefont
  {Shechtman}}, \bibinfo {author} {\bibfnamefont {L.~E.}\ \bibnamefont
  {Weiss}}, \bibinfo {author} {\bibfnamefont {A.~S.}\ \bibnamefont {Backer}},
  \bibinfo {author} {\bibfnamefont {S.~J.}\ \bibnamefont {Sahl}}, \ and\
  \bibinfo {author} {\bibfnamefont {W.~E.}\ \bibnamefont {Moerner}},\
  }\href@noop {} {\bibfield  {journal} {\bibinfo  {journal} {Nano Letters}\
  }\textbf {\bibinfo {volume} {15}},\ \bibinfo {pages} {4194} (\bibinfo {year}
  {2015})}\BibitemShut {NoStop}%
\bibitem [{\citenamefont {Chao}\ \emph {et~al.}(2016)\citenamefont {Chao},
  \citenamefont {Ward},\ and\ \citenamefont
  {Ober}}]{RefWorks:doc:59d667e2e4b04214d8e4df86}%
  \BibitemOpen
  \bibfield  {author} {\bibinfo {author} {\bibfnamefont {J.}~\bibnamefont
  {Chao}}, \bibinfo {author} {\bibfnamefont {E.~S.}\ \bibnamefont {Ward}}, \
  and\ \bibinfo {author} {\bibfnamefont {R.~J.}\ \bibnamefont {Ober}},\
  }\href@noop {} {\bibfield  {journal} {\bibinfo  {journal} {JOSA A}\ }\textbf
  {\bibinfo {volume} {33}},\ \bibinfo {pages} {B57} (\bibinfo {year}
  {2016})}\BibitemShut {NoStop}%
\bibitem [{\citenamefont {Petrov}\ \emph {et~al.}(2017)\citenamefont {Petrov},
  \citenamefont {Shechtman},\ and\ \citenamefont
  {Moerner}}]{RefWorks:doc:59d67d4be4b0dbf6a359f412}%
  \BibitemOpen
  \bibfield  {author} {\bibinfo {author} {\bibfnamefont {P.~N.}\ \bibnamefont
  {Petrov}}, \bibinfo {author} {\bibfnamefont {Y.}~\bibnamefont {Shechtman}}, \
  and\ \bibinfo {author} {\bibfnamefont {W.~E.}\ \bibnamefont {Moerner}},\
  }\href@noop {} {\bibfield  {journal} {\bibinfo  {journal} {Optics Express}\
  }\textbf {\bibinfo {volume} {25}},\ \bibinfo {pages} {7945} (\bibinfo {year}
  {2017})}\BibitemShut {NoStop}%
\bibitem [{\citenamefont {Lew}\ \emph {et~al.}(2013)\citenamefont {Lew},
  \citenamefont {Backlund},\ and\ \citenamefont
  {Moerner}}]{RefWorks:doc:58d6ee0fe4b05fe42c93d968}%
  \BibitemOpen
  \bibfield  {author} {\bibinfo {author} {\bibfnamefont {M.~D.}\ \bibnamefont
  {Lew}}, \bibinfo {author} {\bibfnamefont {M.~P.}\ \bibnamefont {Backlund}}, \
  and\ \bibinfo {author} {\bibfnamefont {W.~E.}\ \bibnamefont {Moerner}},\
  }\href@noop {} {\bibfield  {journal} {\bibinfo  {journal} {Nano Letters}\
  }\textbf {\bibinfo {volume} {13}},\ \bibinfo {pages} {3967} (\bibinfo {year}
  {2013})}\BibitemShut {NoStop}%
\bibitem [{\citenamefont {Backer}\ and\ \citenamefont
  {Moerner}(2014)}]{RefWorks:doc:586d1759e4b05e0ea0634282}%
  \BibitemOpen
  \bibfield  {author} {\bibinfo {author} {\bibfnamefont {A.~S.}\ \bibnamefont
  {Backer}}\ and\ \bibinfo {author} {\bibfnamefont {W.~E.}\ \bibnamefont
  {Moerner}},\ }\href@noop {} {\bibfield  {journal} {\bibinfo  {journal} {The
  Journal of Physical Chemistry B}\ }\textbf {\bibinfo {volume} {118}},\
  \bibinfo {pages} {8313} (\bibinfo {year} {2014})}\BibitemShut {NoStop}%
\bibitem [{\citenamefont {Pezz�}\ \emph {et~al.}(2017)\citenamefont
  {Pezz�}, \citenamefont {Ciampini}, \citenamefont {Spagnolo}, \citenamefont
  {Humphreys}, \citenamefont {Datta}, \citenamefont {Walmsley}, \citenamefont
  {Barbieri}, \citenamefont {Sciarrino},\ and\ \citenamefont
  {Smerzi}}]{RefWorks:doc:5a96bc0de4b025568d7175de}%
  \BibitemOpen
  \bibfield  {author} {\bibinfo {author} {\bibfnamefont {L.}~\bibnamefont
  {Pezz�}}, \bibinfo {author} {\bibfnamefont {M.~A.}\ \bibnamefont
  {Ciampini}}, \bibinfo {author} {\bibfnamefont {N.}~\bibnamefont {Spagnolo}},
  \bibinfo {author} {\bibfnamefont {P.~C.}\ \bibnamefont {Humphreys}}, \bibinfo
  {author} {\bibfnamefont {A.}~\bibnamefont {Datta}}, \bibinfo {author}
  {\bibfnamefont {I.~A.}\ \bibnamefont {Walmsley}}, \bibinfo {author}
  {\bibfnamefont {M.}~\bibnamefont {Barbieri}}, \bibinfo {author}
  {\bibfnamefont {F.}~\bibnamefont {Sciarrino}}, \ and\ \bibinfo {author}
  {\bibfnamefont {A.}~\bibnamefont {Smerzi}},\ }\href@noop {} {\bibfield
  {journal} {\bibinfo  {journal} {Physical Review Letters}\ }\textbf {\bibinfo
  {volume} {119}},\ \bibinfo {pages} {130504} (\bibinfo {year}
  {2017})}\BibitemShut {NoStop}%
\bibitem [{\citenamefont {Ciampini}\ \emph {et~al.}(2016)\citenamefont
  {Ciampini}, \citenamefont {Spagnolo}, \citenamefont {Vitelli}, \citenamefont
  {Pezz}, \citenamefont {Smerzi},\ and\ \citenamefont
  {Sciarrino}}]{RefWorks:doc:59db8849e4b04214d8e577ff}%
  \BibitemOpen
  \bibfield  {author} {\bibinfo {author} {\bibfnamefont {M.~A.}\ \bibnamefont
  {Ciampini}}, \bibinfo {author} {\bibfnamefont {N.}~\bibnamefont {Spagnolo}},
  \bibinfo {author} {\bibfnamefont {C.}~\bibnamefont {Vitelli}}, \bibinfo
  {author} {\bibfnamefont {L.}~\bibnamefont {Pezz}}, \bibinfo {author}
  {\bibfnamefont {A.}~\bibnamefont {Smerzi}}, \ and\ \bibinfo {author}
  {\bibfnamefont {F.}~\bibnamefont {Sciarrino}},\ }\href@noop {} {\bibfield
  {journal} {\bibinfo  {journal} {Scientific Reports}\ }\textbf {\bibinfo
  {volume} {6}},\ \bibinfo {pages} {28881} (\bibinfo {year}
  {2016})}\BibitemShut {NoStop}%
\bibitem [{\citenamefont
  {Goodman}(2005)}]{RefWorks:doc:58b4a33ce4b07d79a1c92f46}%
  \BibitemOpen
  \bibfield  {author} {\bibinfo {author} {\bibfnamefont {J.~W.}\ \bibnamefont
  {Goodman}},\ }\href@noop {} {\emph {\bibinfo {title} {Introduction to Fourier
  Optics}}}\ (\bibinfo  {publisher} {Roberts and Company Publishers},\ \bibinfo
  {year} {2005})\BibitemShut {NoStop}%
\bibitem [{Ref()}]{RefWorks:doc:59d6a369e4b04214d8e4ea74}%
  \BibitemOpen
  \href@noop {} {\bibinfo  {journal} {Supplemental Material}\ }\BibitemShut
  {NoStop}%
\bibitem [{\citenamefont
  {Helstrom}(1967)}]{RefWorks:doc:59d68c3ce4b03cc71b452e19}%
  \BibitemOpen
\bibfield  {journal} {  }\bibfield  {author} {\bibinfo {author} {\bibfnamefont
  {C.~W.}\ \bibnamefont {Helstrom}},\ }\href@noop {} {\bibfield  {journal}
  {\bibinfo  {journal} {Physics Letters A}\ }\textbf {\bibinfo {volume} {25}},\
  \bibinfo {pages} {101} (\bibinfo {year} {1967})}\BibitemShut {NoStop}%
\bibitem [{\citenamefont
  {Helstrom}(1970)}]{RefWorks:doc:59d68c70e4b04e4ae184ba67}%
  \BibitemOpen
  \bibfield  {author} {\bibinfo {author} {\bibfnamefont {C.~W.}\ \bibnamefont
  {Helstrom}},\ }\href@noop {} {\bibfield  {journal} {\bibinfo  {journal}
  {JOSA}\ }\textbf {\bibinfo {volume} {60}},\ \bibinfo {pages} {233} (\bibinfo
  {year} {1970})}\BibitemShut {NoStop}%
\bibitem [{\citenamefont
  {Holevo}(2011)}]{RefWorks:doc:59d6a92ae4b04e4ae184bf2b}%
  \BibitemOpen
  \bibfield  {author} {\bibinfo {author} {\bibfnamefont {A.~S.}\ \bibnamefont
  {Holevo}},\ }\href@noop {} {\emph {\bibinfo {title} {Probabilistic and
  Statistical Aspects of Quantum Theory}}},\ Vol.~\bibinfo {volume} {1}\
  (\bibinfo  {publisher} {Springer Science and Business Media},\ \bibinfo
  {year} {2011})\BibitemShut {NoStop}%
\bibitem [{\citenamefont {Braunstein}\ and\ \citenamefont
  {Caves}(1994)}]{RefWorks:doc:59d6a9afe4b03cc71b45322e}%
  \BibitemOpen
  \bibfield  {author} {\bibinfo {author} {\bibfnamefont {S.~L.}\ \bibnamefont
  {Braunstein}}\ and\ \bibinfo {author} {\bibfnamefont {C.~M.}\ \bibnamefont
  {Caves}},\ }\href@noop {} {\bibfield  {journal} {\bibinfo  {journal}
  {Physical Review Letters}\ }\textbf {\bibinfo {volume} {72}},\ \bibinfo
  {pages} {3439} (\bibinfo {year} {1994})}\BibitemShut {NoStop}%
\bibitem [{\citenamefont {Rehacek}\ \emph
  {et~al.}(2017{\natexlab{a}})\citenamefont {Rehacek}, \citenamefont {Paur},
  \citenamefont {Stoklasa}, \citenamefont {Hradil},\ and\ \citenamefont
  {Sanchez-Soto}}]{RefWorks:doc:59d788d0e4b0573d0abb92f2}%
  \BibitemOpen
  \bibfield  {author} {\bibinfo {author} {\bibfnamefont {J.}~\bibnamefont
  {Rehacek}}, \bibinfo {author} {\bibfnamefont {M.}~\bibnamefont {Paur}},
  \bibinfo {author} {\bibfnamefont {B.}~\bibnamefont {Stoklasa}}, \bibinfo
  {author} {\bibfnamefont {Z.}~\bibnamefont {Hradil}}, \ and\ \bibinfo {author}
  {\bibfnamefont {L.~L.}\ \bibnamefont {Sanchez-Soto}},\ }\href@noop {}
  {\bibfield  {journal} {\bibinfo  {journal} {Optics Letters}\ }\textbf
  {\bibinfo {volume} {42}},\ \bibinfo {pages} {231} (\bibinfo {year}
  {2017}{\natexlab{a}})}\BibitemShut {NoStop}%
\bibitem [{\citenamefont {Paur}\ \emph {et~al.}(2016)\citenamefont {Paur},
  \citenamefont {Stoklasa}, \citenamefont {Hradil}, \citenamefont
  {Sanchez-Soto},\ and\ \citenamefont
  {Rehacek}}]{RefWorks:doc:59d788a4e4b03cc71b455654}%
  \BibitemOpen
  \bibfield  {author} {\bibinfo {author} {\bibfnamefont {M.}~\bibnamefont
  {Paur}}, \bibinfo {author} {\bibfnamefont {B.}~\bibnamefont {Stoklasa}},
  \bibinfo {author} {\bibfnamefont {Z.}~\bibnamefont {Hradil}}, \bibinfo
  {author} {\bibfnamefont {L.~L.}\ \bibnamefont {Sanchez-Soto}}, \ and\
  \bibinfo {author} {\bibfnamefont {J.}~\bibnamefont {Rehacek}},\ }\href@noop
  {} {\bibfield  {journal} {\bibinfo  {journal} {Optica}\ }\textbf {\bibinfo
  {volume} {3}},\ \bibinfo {pages} {1144} (\bibinfo {year} {2016})}\BibitemShut
  {NoStop}%
\bibitem [{\citenamefont {Tang}\ \emph {et~al.}(2016)\citenamefont {Tang},
  \citenamefont {Durak},\ and\ \citenamefont
  {Ling}}]{RefWorks:doc:59d78880e4b03cc71b45564f}%
  \BibitemOpen
  \bibfield  {author} {\bibinfo {author} {\bibfnamefont {Z.~S.}\ \bibnamefont
  {Tang}}, \bibinfo {author} {\bibfnamefont {K.}~\bibnamefont {Durak}}, \ and\
  \bibinfo {author} {\bibfnamefont {A.}~\bibnamefont {Ling}},\ }\href@noop {}
  {\bibfield  {journal} {\bibinfo  {journal} {Optics Express}\ }\textbf
  {\bibinfo {volume} {24}},\ \bibinfo {pages} {22004} (\bibinfo {year}
  {2016})}\BibitemShut {NoStop}%
\bibitem [{\citenamefont {Tham}\ \emph {et~al.}(2017)\citenamefont {Tham},
  \citenamefont {Ferretti},\ and\ \citenamefont
  {Steinberg}}]{RefWorks:doc:59d78d46e4b0573d0abb9352}%
  \BibitemOpen
  \bibfield  {author} {\bibinfo {author} {\bibfnamefont {W.-K.}\ \bibnamefont
  {Tham}}, \bibinfo {author} {\bibfnamefont {H.}~\bibnamefont {Ferretti}}, \
  and\ \bibinfo {author} {\bibfnamefont {A.~M.}\ \bibnamefont {Steinberg}},\
  }\href@noop {} {\bibfield  {journal} {\bibinfo  {journal} {Physical Review
  Letters}\ }\textbf {\bibinfo {volume} {118}},\ \bibinfo {pages} {070801}
  (\bibinfo {year} {2017})}\BibitemShut {NoStop}%
\bibitem [{\citenamefont {Yang}\ \emph {et~al.}(2016)\citenamefont {Yang},
  \citenamefont {Tashchilina}, \citenamefont {Moiseev}, \citenamefont {Simon},\
  and\ \citenamefont {Lvovsky}}]{RefWorks:doc:59d78833e4b00f3d39178ac9}%
  \BibitemOpen
  \bibfield  {author} {\bibinfo {author} {\bibfnamefont {F.}~\bibnamefont
  {Yang}}, \bibinfo {author} {\bibfnamefont {A.}~\bibnamefont {Tashchilina}},
  \bibinfo {author} {\bibfnamefont {E.~S.}\ \bibnamefont {Moiseev}}, \bibinfo
  {author} {\bibfnamefont {C.}~\bibnamefont {Simon}}, \ and\ \bibinfo {author}
  {\bibfnamefont {A.~I.}\ \bibnamefont {Lvovsky}},\ }\href@noop {} {\bibfield
  {journal} {\bibinfo  {journal} {Optica}\ }\textbf {\bibinfo {volume} {3}},\
  \bibinfo {pages} {1148} (\bibinfo {year} {2016})}\BibitemShut {NoStop}%
\bibitem [{\citenamefont {Nair}\ and\ \citenamefont
  {Tsang}(2016{\natexlab{a}})}]{RefWorks:doc:59d786c1e4b061a5d614b825}%
  \BibitemOpen
  \bibfield  {author} {\bibinfo {author} {\bibfnamefont {R.}~\bibnamefont
  {Nair}}\ and\ \bibinfo {author} {\bibfnamefont {M.}~\bibnamefont {Tsang}},\
  }\href@noop {} {\bibfield  {journal} {\bibinfo  {journal} {Physical Review
  Letters}\ }\textbf {\bibinfo {volume} {117}},\ \bibinfo {pages} {190801}
  (\bibinfo {year} {2016}{\natexlab{a}})}\BibitemShut {NoStop}%
\bibitem [{\citenamefont {Lupo}\ and\ \citenamefont
  {Pirandola}(2016)}]{RefWorks:doc:59d78724e4b0dbf6a35a2153}%
  \BibitemOpen
  \bibfield  {author} {\bibinfo {author} {\bibfnamefont {C.}~\bibnamefont
  {Lupo}}\ and\ \bibinfo {author} {\bibfnamefont {S.}~\bibnamefont
  {Pirandola}},\ }\href@noop {} {\bibfield  {journal} {\bibinfo  {journal}
  {Physical Review Letters}\ }\textbf {\bibinfo {volume} {117}},\ \bibinfo
  {pages} {190802} (\bibinfo {year} {2016})}\BibitemShut {NoStop}%
\bibitem [{\citenamefont {Lu}\ \emph {et~al.}(2016)\citenamefont {Lu},
  \citenamefont {Nair},\ and\ \citenamefont
  {Tsang}}]{RefWorks:doc:59d78987e4b061a5d614b87d}%
  \BibitemOpen
  \bibfield  {author} {\bibinfo {author} {\bibfnamefont {X.-M.}\ \bibnamefont
  {Lu}}, \bibinfo {author} {\bibfnamefont {R.}~\bibnamefont {Nair}}, \ and\
  \bibinfo {author} {\bibfnamefont {M.}~\bibnamefont {Tsang}},\ }\href@noop {}
  {\bibfield  {journal} {\bibinfo  {journal} {arXiv preprint arXiv:1609.03025}\
  } (\bibinfo {year} {2016})}\BibitemShut {NoStop}%
\bibitem [{\citenamefont {Kerviche}\ \emph {et~al.}(2017)\citenamefont
  {Kerviche}, \citenamefont {Guha},\ and\ \citenamefont
  {Ashok}}]{RefWorks:doc:59d78a88e4b00aed3eaf7730}%
  \BibitemOpen
  \bibfield  {author} {\bibinfo {author} {\bibfnamefont {R.}~\bibnamefont
  {Kerviche}}, \bibinfo {author} {\bibfnamefont {S.}~\bibnamefont {Guha}}, \
  and\ \bibinfo {author} {\bibfnamefont {A.}~\bibnamefont {Ashok}},\
  }\href@noop {} {\bibfield  {journal} {\bibinfo  {journal} {arXiv preprint
  arXiv:1701.04913}\ } (\bibinfo {year} {2017})}\BibitemShut {NoStop}%
\bibitem [{\citenamefont {Rehacek}\ \emph
  {et~al.}(2017{\natexlab{b}})\citenamefont {Rehacek}, \citenamefont {Hradil},
  \citenamefont {Stoklasa}, \citenamefont {Paur}, \citenamefont {Grover},
  \citenamefont {Krzic},\ and\ \citenamefont
  {Sanchez-Soto}}]{RefWorks:doc:5a96bb34e4b01d55d0b35be4}%
  \BibitemOpen
  \bibfield  {author} {\bibinfo {author} {\bibfnamefont {J.}~\bibnamefont
  {Rehacek}}, \bibinfo {author} {\bibfnamefont {Z.}~\bibnamefont {Hradil}},
  \bibinfo {author} {\bibfnamefont {B.}~\bibnamefont {Stoklasa}}, \bibinfo
  {author} {\bibfnamefont {M.}~\bibnamefont {Paur}}, \bibinfo {author}
  {\bibfnamefont {J.}~\bibnamefont {Grover}}, \bibinfo {author} {\bibfnamefont
  {A.}~\bibnamefont {Krzic}}, \ and\ \bibinfo {author} {\bibfnamefont {L.~L.}\
  \bibnamefont {Sanchez-Soto}},\ }\href@noop {} {\bibfield  {journal} {\bibinfo
   {journal} {Physical Review A}\ }\textbf {\bibinfo {volume} {96}},\ \bibinfo
  {pages} {062107} (\bibinfo {year} {2017}{\natexlab{b}})}\BibitemShut
  {NoStop}%
\bibitem [{\citenamefont {Nair}\ and\ \citenamefont
  {Tsang}(2016{\natexlab{b}})}]{RefWorks:doc:59d78afee4b0dbf6a35a21ab}%
  \BibitemOpen
  \bibfield  {author} {\bibinfo {author} {\bibfnamefont {R.}~\bibnamefont
  {Nair}}\ and\ \bibinfo {author} {\bibfnamefont {M.}~\bibnamefont {Tsang}},\
  }\href@noop {} {\bibfield  {journal} {\bibinfo  {journal} {Optics Express}\
  }\textbf {\bibinfo {volume} {24}},\ \bibinfo {pages} {3684} (\bibinfo {year}
  {2016}{\natexlab{b}})}\BibitemShut {NoStop}%
\bibitem [{\citenamefont {Ang}\ \emph {et~al.}(2017)\citenamefont {Ang},
  \citenamefont {Nair},\ and\ \citenamefont
  {Tsang}}]{RefWorks:doc:59d78b2fe4b00aed3eaf773f}%
  \BibitemOpen
  \bibfield  {author} {\bibinfo {author} {\bibfnamefont {S.~Z.}\ \bibnamefont
  {Ang}}, \bibinfo {author} {\bibfnamefont {R.}~\bibnamefont {Nair}}, \ and\
  \bibinfo {author} {\bibfnamefont {M.}~\bibnamefont {Tsang}},\ }\href@noop {}
  {\bibfield  {journal} {\bibinfo  {journal} {Physical Review A}\ }\textbf
  {\bibinfo {volume} {95}},\ \bibinfo {pages} {063847} (\bibinfo {year}
  {2017})}\BibitemShut {NoStop}%
\bibitem [{\citenamefont
  {Tsang}(2015)}]{RefWorks:doc:59d79217e4b04e4ae184e15d}%
  \BibitemOpen
  \bibfield  {author} {\bibinfo {author} {\bibfnamefont {M.}~\bibnamefont
  {Tsang}},\ }\href@noop {} {\bibfield  {journal} {\bibinfo  {journal}
  {Optica}\ }\textbf {\bibinfo {volume} {2}},\ \bibinfo {pages} {646} (\bibinfo
  {year} {2015})}\BibitemShut {NoStop}%
\bibitem [{\citenamefont {Fujiwara}\ and\ \citenamefont
  {Nagaoka}(1995)}]{RefWorks:doc:59db836ce4b061a5d61513ae}%
  \BibitemOpen
  \bibfield  {author} {\bibinfo {author} {\bibfnamefont {A.}~\bibnamefont
  {Fujiwara}}\ and\ \bibinfo {author} {\bibfnamefont {H.}~\bibnamefont
  {Nagaoka}},\ }\href@noop {} {\bibfield  {journal} {\bibinfo  {journal}
  {Physics Letters A}\ }\textbf {\bibinfo {volume} {201}},\ \bibinfo {pages}
  {119} (\bibinfo {year} {1995})}\BibitemShut {NoStop}%
\bibitem [{\citenamefont {Gill}\ and\ \citenamefont
  {Massar}(2000)}]{RefWorks:doc:59db838de4b04e4ae1854dc1}%
  \BibitemOpen
  \bibfield  {author} {\bibinfo {author} {\bibfnamefont {R.~D.}\ \bibnamefont
  {Gill}}\ and\ \bibinfo {author} {\bibfnamefont {S.}~\bibnamefont {Massar}},\
  }\href@noop {} {\bibfield  {journal} {\bibinfo  {journal} {Physical Review
  A}\ }\textbf {\bibinfo {volume} {61}},\ \bibinfo {pages} {042312} (\bibinfo
  {year} {2000})}\BibitemShut {NoStop}%
\bibitem [{\citenamefont {Barndorff-Nielsen}\ and\ \citenamefont
  {Gill}(2000)}]{RefWorks:doc:59db88ece4b00f3d3917fe68}%
  \BibitemOpen
  \bibfield  {author} {\bibinfo {author} {\bibfnamefont {O.~E.}\ \bibnamefont
  {Barndorff-Nielsen}}\ and\ \bibinfo {author} {\bibfnamefont {R.~D.}\
  \bibnamefont {Gill}},\ }\href@noop {} {\bibfield  {journal} {\bibinfo
  {journal} {Journal of Physics A: Mathematical and General}\ }\textbf
  {\bibinfo {volume} {33}},\ \bibinfo {pages} {4481} (\bibinfo {year}
  {2000})}\BibitemShut {NoStop}%
\bibitem [{\citenamefont
  {Fujiwara}(2006)}]{RefWorks:doc:59db88c9e4b03cc71b45d83a}%
  \BibitemOpen
  \bibfield  {author} {\bibinfo {author} {\bibfnamefont {A.}~\bibnamefont
  {Fujiwara}},\ }\href@noop {} {\bibfield  {journal} {\bibinfo  {journal}
  {Journal of Physics A: Mathematical and General}\ }\textbf {\bibinfo {volume}
  {39}},\ \bibinfo {pages} {12489} (\bibinfo {year} {2006})}\BibitemShut
  {NoStop}%
\bibitem [{\citenamefont
  {Luati}(2004)}]{RefWorks:doc:59db8892e4b04214d8e57811}%
  \BibitemOpen
  \bibfield  {author} {\bibinfo {author} {\bibfnamefont {A.}~\bibnamefont
  {Luati}},\ }\href@noop {} {\bibfield  {journal} {\bibinfo  {journal} {The
  Annals of Statistics}\ }\textbf {\bibinfo {volume} {32}},\ \bibinfo {pages}
  {1770} (\bibinfo {year} {2004})}\BibitemShut {NoStop}%
\bibitem [{\citenamefont
  {Matsumoto}(2002)}]{RefWorks:doc:59db87e9e4b00f3d3917fe45}%
  \BibitemOpen
  \bibfield  {author} {\bibinfo {author} {\bibfnamefont {K.}~\bibnamefont
  {Matsumoto}},\ }\href@noop {} {\bibfield  {journal} {\bibinfo  {journal}
  {Journal of Physics A: Mathematical and General}\ }\textbf {\bibinfo {volume}
  {35}},\ \bibinfo {pages} {3111} (\bibinfo {year} {2002})}\BibitemShut
  {NoStop}%
\bibitem [{\citenamefont {Humphreys}\ \emph {et~al.}(2013)\citenamefont
  {Humphreys}, \citenamefont {Barbieri}, \citenamefont {Datta},\ and\
  \citenamefont {Walmsley}}]{RefWorks:doc:59db8809e4b00f3d3917fe4a}%
  \BibitemOpen
  \bibfield  {author} {\bibinfo {author} {\bibfnamefont {P.~C.}\ \bibnamefont
  {Humphreys}}, \bibinfo {author} {\bibfnamefont {M.}~\bibnamefont {Barbieri}},
  \bibinfo {author} {\bibfnamefont {A.}~\bibnamefont {Datta}}, \ and\ \bibinfo
  {author} {\bibfnamefont {I.~A.}\ \bibnamefont {Walmsley}},\ }\href@noop {}
  {\bibfield  {journal} {\bibinfo  {journal} {Physical Review Letters}\
  }\textbf {\bibinfo {volume} {111}},\ \bibinfo {pages} {070403} (\bibinfo
  {year} {2013})}\BibitemShut {NoStop}%
\bibitem [{\citenamefont {Chen}\ and\ \citenamefont
  {Yuan}(2017)}]{RefWorks:doc:59db886de4b0dbf6a35a90d3}%
  \BibitemOpen
  \bibfield  {author} {\bibinfo {author} {\bibfnamefont {Y.}~\bibnamefont
  {Chen}}\ and\ \bibinfo {author} {\bibfnamefont {H.}~\bibnamefont {Yuan}},\
  }\href@noop {} {\bibfield  {journal} {\bibinfo  {journal} {New Journal of
  Physics}\ } (\bibinfo {year} {2017})}\BibitemShut {NoStop}%
\bibitem [{\citenamefont
  {Bryngdahl}(1971)}]{RefWorks:doc:59db8f6ee4b04214d8e57987}%
  \BibitemOpen
  \bibfield  {author} {\bibinfo {author} {\bibfnamefont {O.}~\bibnamefont
  {Bryngdahl}},\ }\href@noop {} {\bibfield  {journal} {\bibinfo  {journal}
  {JOSA}\ }\textbf {\bibinfo {volume} {61}},\ \bibinfo {pages} {169} (\bibinfo
  {year} {1971})}\BibitemShut {NoStop}%
\bibitem [{\citenamefont {Hohlbein}\ and\ \citenamefont
  {Hubner}(2005)}]{RefWorks:doc:59db9222e4b0573d0abc062b}%
  \BibitemOpen
  \bibfield  {author} {\bibinfo {author} {\bibfnamefont {J.}~\bibnamefont
  {Hohlbein}}\ and\ \bibinfo {author} {\bibfnamefont {C.~G.}\ \bibnamefont
  {Hubner}},\ }\href@noop {} {\bibfield  {journal} {\bibinfo  {journal}
  {Applied Physics Letters}\ }\textbf {\bibinfo {volume} {86}},\ \bibinfo
  {pages} {121104} (\bibinfo {year} {2005})}\BibitemShut {NoStop}%
\bibitem [{\citenamefont {Jarutis}\ \emph {et~al.}(2000)\citenamefont
  {Jarutis}, \citenamefont {Paškauskas},\ and\ \citenamefont
  {Stabinis}}]{RefWorks:doc:59db8fc9e4b00aed3eaff1ab}%
  \BibitemOpen
  \bibfield  {author} {\bibinfo {author} {\bibfnamefont {V.}~\bibnamefont
  {Jarutis}}, \bibinfo {author} {\bibfnamefont {R.}~\bibnamefont
  {Paškauskas}}, \ and\ \bibinfo {author} {\bibfnamefont {A.}~\bibnamefont
  {Stabinis}},\ }\href@noop {} {\bibfield  {journal} {\bibinfo  {journal}
  {Optics Communications}\ }\textbf {\bibinfo {volume} {184}},\ \bibinfo
  {pages} {105} (\bibinfo {year} {2000})}\BibitemShut {NoStop}%
\bibitem [{\citenamefont {Wang}\ \emph {et~al.}(2017)\citenamefont {Wang},
  \citenamefont {Yan}, \citenamefont {Friberg}, \citenamefont {Kuebel},\ and\
  \citenamefont {Visser}}]{RefWorks:doc:59db8fefe4b03cc71b45d9cf}%
  \BibitemOpen
  \bibfield  {author} {\bibinfo {author} {\bibfnamefont {Y.}~\bibnamefont
  {Wang}}, \bibinfo {author} {\bibfnamefont {S.}~\bibnamefont {Yan}}, \bibinfo
  {author} {\bibfnamefont {A.~T.}\ \bibnamefont {Friberg}}, \bibinfo {author}
  {\bibfnamefont {D.}~\bibnamefont {Kuebel}}, \ and\ \bibinfo {author}
  {\bibfnamefont {T.~D.}\ \bibnamefont {Visser}},\ }\href@noop {} {\bibfield
  {journal} {\bibinfo  {journal} {JOSA A}\ }\textbf {\bibinfo {volume} {34}},\
  \bibinfo {pages} {1201} (\bibinfo {year} {2017})}\BibitemShut {NoStop}%
\bibitem [{\citenamefont
  {Shapiro}(2009)}]{RefWorks:doc:5a15b8c4e4b0c5584aef9f41}%
  \BibitemOpen
  \bibfield  {author} {\bibinfo {author} {\bibfnamefont {J.~H.}\ \bibnamefont
  {Shapiro}},\ }\href@noop {} {\bibfield  {journal} {\bibinfo  {journal} {IEEE
  Journal of Selected Topics in Quantum Electronics}\ }\textbf {\bibinfo
  {volume} {15}},\ \bibinfo {pages} {1547} (\bibinfo {year}
  {2009})}\BibitemShut {NoStop}%
\bibitem [{\citenamefont {Balzarotti}\ \emph {et~al.}(2016)\citenamefont
  {Balzarotti}, \citenamefont {Eilers}, \citenamefont {Gwosch}, \citenamefont
  {Gynn}, \citenamefont {Westphal}, \citenamefont {Stefani}, \citenamefont
  {Elf},\ and\ \citenamefont {Hell}}]{RefWorks:doc:59f64286e4b028f0a93ad056}%
  \BibitemOpen
  \bibfield  {author} {\bibinfo {author} {\bibfnamefont {F.}~\bibnamefont
  {Balzarotti}}, \bibinfo {author} {\bibfnamefont {Y.}~\bibnamefont {Eilers}},
  \bibinfo {author} {\bibfnamefont {K.~C.}\ \bibnamefont {Gwosch}}, \bibinfo
  {author} {\bibfnamefont {A.~H.}\ \bibnamefont {Gynn}}, \bibinfo {author}
  {\bibfnamefont {V.}~\bibnamefont {Westphal}}, \bibinfo {author}
  {\bibfnamefont {F.~D.}\ \bibnamefont {Stefani}}, \bibinfo {author}
  {\bibfnamefont {J.}~\bibnamefont {Elf}}, \ and\ \bibinfo {author}
  {\bibfnamefont {S.~W.}\ \bibnamefont {Hell}},\ }\href@noop {} {\bibfield
  {journal} {\bibinfo  {journal} {Science}\ ,\ \bibinfo {pages} {aak9913}}
  (\bibinfo {year} {2016})}\BibitemShut {NoStop}%
\bibitem [{\citenamefont {Backlund}\ \emph {et~al.}(2014)\citenamefont
  {Backlund}, \citenamefont {Lew}, \citenamefont {Backer}, \citenamefont
  {Sahl},\ and\ \citenamefont
  {Moerner}}]{RefWorks:doc:58d6edb6e4b0b1f71a1fc890}%
  \BibitemOpen
  \bibfield  {author} {\bibinfo {author} {\bibfnamefont {M.~P.}\ \bibnamefont
  {Backlund}}, \bibinfo {author} {\bibfnamefont {M.~D.}\ \bibnamefont {Lew}},
  \bibinfo {author} {\bibfnamefont {A.~S.}\ \bibnamefont {Backer}}, \bibinfo
  {author} {\bibfnamefont {S.~J.}\ \bibnamefont {Sahl}}, \ and\ \bibinfo
  {author} {\bibfnamefont {W.~E.}\ \bibnamefont {Moerner}},\ }\href@noop {}
  {\bibfield  {journal} {\bibinfo  {journal} {ChemPhysChem}\ }\textbf {\bibinfo
  {volume} {15}},\ \bibinfo {pages} {587} (\bibinfo {year} {2014})}\BibitemShut
  {NoStop}%
\end{thebibliography}%
\end{document}